\documentclass[a4paper,11 pt]{article}
\pdfoutput=1 

\usepackage{jheppub} 

\usepackage[T1]{fontenc}
\usepackage{amsmath}
\usepackage{cancel}
\usepackage{caption}
\usepackage{subcaption,longtable,stmaryrd}
\usepackage{comment}

\DeclareSymbolFont{usualmathcal}{OMS}{cmsy}{m}{n}
\DeclareSymbolFontAlphabet{\mathcal}{usualmathcal}

\usepackage{bigints}
\usepackage{multicol}
\usepackage{blindtext}
\usepackage{tikz}
\usepackage{enumitem}
\usepackage{amsmath, amssymb, graphics, setspace}
\newcommand{\mathsym}[1]{{}}
\renewcommand{\L}{{\mathcal{L}}}
\newcommand{\bL}{\bar{{\mathcal{L}}}}
\newcommand{\z}{{\bar z}}
\newcommand{\h}{{\bar h}}
\newcommand{\unicode}[1]{{}}
\usepackage[mathscr]{euscript}
\DeclareMathOperator\erf{erf}
\DeclareSymbolFont{rsfs}{U}{rsfs}{m}{n}
\DeclareSymbolFontAlphabet{\mathscrsfs}{rsfs}
\DeclareSymbolFont{rmlargesymbols}{OMX}{mdbch}{m}{n}
\DeclareMathSymbol{\rmintop}{\mathop}{rmlargesymbols}{82}
\DeclareMathSymbol{\rmointop}{\mathop}{rmlargesymbols}{72}
\usepackage[customcolors,shade]{hf-tikz}

\usepackage{graphicx}

\definecolor{darkbrown}{rgb}{0.787, 0.26, 0.187}
\definecolor{alizarin}{rgb}{0.82, 0.1, 0.26}

\title{Symmetry Resolution in non-Lorentzian Field Theories}

\author[a]{Aritra Banerjee,}
\author[b,c]{Rudranil Basu,}
\author[d]{Arpan Bhattacharyya,}
\author[d]{and Nilachal Chakrabarti}
\affiliation[a]{Birla Institute of Technology and Science, Pilani Campus, Rajasthan 333031, India.}
\affiliation[b]{Department of Physics, Birla Institute of Technology and Science Pilani, Zuarinagar, Goa 403726, India.
}
\affiliation[c]{Center for Research in Quantum Information and Technology,
Birla Institute of Technology and Science Pilani, Zuarinagar, Goa 403726, India.}
\affiliation[d]{Department of Physics, Indian Institute of Technology Gandhinagar, Gujarat 382055, India.}

\emailAdd{aritra.banerjee@pilani.bits-pilani.ac.in}
\emailAdd{rudranilb@goa.bits-pilani.ac.in}
\emailAdd{abhattacharyya@iitgn.ac.in}
\emailAdd{nilachalchakrabarti@iitgn.ac.in}

\abstract{Starting from the computation of Symmetry Resolved Entanglement Entropy (SREE) for boosted intervals in a two dimensional Conformal Field Theory, we compute the same in various non-Lorentzian limits, viz, Galilean and Carrollian Conformal Field Theory in same number of dimensions. We approach the problem both from a limiting perspective and by using intrinsic symmetries of respective non-Lorentzian conformal algebras. In particular, we calculate the leading order terms, logarithmic terms, and the $\mathcal{O}(1)$ terms and explicitly show exact compliance with \textit{equipartition of entanglement}, even in the non-Lorentzian system. Keeping in mind the holographic origin of SREE for the Carrollian limit, we further compute SREE for BMS$_{3}$-Kac-Moody, which couples a $U(1)\times U(1)$ theory with bulk gravity.}

\makeatletter

\begin{document}

\maketitle

\section{Introduction}
Over the past few decades, the concept of entanglement has played an important role in understanding several important aspects of physical systems. It gives us pertinent information about the non-local correlation between quantum systems \cite{Horodecki:2009zz}. One of the common and widely used measures of entanglement (for the pure state) is the von Neumann or entanglement entropy (EE) \cite{bennett}, which has found important applications in diverse branches of physics, e.g. quantum many-body system, quantum computation, black hole physics, as well as in the context of holography (more specifically AdS/CFT and other cousins thereof) \cite{Horodecki:2009zz,Amico:2007ag,Laflorencie:2015eck,Rangamani:2016dms,Almheiri:2020cfm,Raju:2020smc,Bhattacharyya:2015nvf}. Several interesting results for EE  have emerged in the recent past, particularly for $1+1$d quantum field theories (QFTs) and conformal field theories (CFTs) \cite{Bombelli:1986rw, Srednicki:1993im,Calabrese:2004eu,Calabrese:2009qy}. Analytical (for CFTs) and numerical studies (for certain spin chain models in $1+1$d, as well as lattice QFT models) have established that EE shows universal scaling at the critical point \cite{Bombelli:1986rw, Srednicki:1993im,Jin,Latorre2,Casini:2006hu,Casini:2009sr,Casini:2013rba}. Since this is where the underlying system can be effectively described by a CFT, it inevitably makes EE a useful tool for diagnosing phase transitions and critical points for quantum many-body systems. Efficient algorithms, such as tensor network methods \cite{White:1992zz,Scho,Verstraete:2008cex,Vidal:2008zz,Bridgeman:2016dhh}, have been developed utilizing the underlying entanglement structure to simulate the ground states of quantum many-body systems \footnote{Tensor networks also have found interesting applications in the context of AdS/CFT correspondence, see for example: \cite{Vidal:2008zz,TenHol1,TenHol2,TenHol3,TenHol4,TenHol5,TenHol6}.} . In addition to EE, considerable progress has been made in computing Renyi entropies (RE) \cite{renyi}, particularly for lower-dimensional quantum systems \cite{Franchini:2007eu,renyi1,renyi2,renyi3,Klebanov:2011uf}. This has helped us access the full entanglement spectrum of the underlying system. Although there has been considerable progress in the theoretical understanding of EE and RE for various interesting quantum systems, only recently  there has been some progress in experimentally accessing these quantities. As a notable example, the second RE has been extracted from a bosonic cold atomic system \cite{Islam}. 
\medskip

Not only EE, but there is a plethora of other measures of entanglement. EE is a good measure of entanglement, but only for a pure state. The choice of an appropriate measure of entanglement depends on various factors, e.g. purity of the underlying state, temperature, and how partitioning of the system is done \cite{Plenio}. Usually, a good measure of entanglement is expected to satisfy certain properties; e.g., it should be zero for a separable state and decrease monotonically under LOCC (local operation and classical communication). Based on these, several choices for measures of entanglement have been proposed, e.g., entanglement of formation \cite{Bennett:1996gf}, relative entropy of entanglement \cite{Vedral:1997qn}, squashed entanglement \cite{Tucci:2002gzf,squash}, negativity \cite{Vidal:2002zz,Calabrese:2012ew} \footnote{Note that negativity does not satisfy all properties required for entanglement measures. Also, it does not coincide with the EE (or the von Neumann entropy) when the total system is pure. In fact it can be zero even for certain entangled states \cite{Rains:1998gp}.}  etc, and some progress has been made in computing them for certain simple QFTs and quantum many-body systems \cite{Calabrese:2012ew,Calabrese:2014yza,Bhattacharyya:2018sbw,Bhattacharyya:2019tsi,Mollabashi:2020yie}. All these measures are typically functions of the eigenvalues of the reduced density matrix of the subsystem of interest.
\medskip

In particular, one needs to generally diagonalize the reduced density matrix for the computation of EE. Now, a natural question arises: What happens to the structure of the reduced density matrix in the presence of symmetry? As the reduced density matrix will commute with symmetry operators acting on the subsystem, it will take a block diagonal form where each block corresponds to different symmetry sectors \cite{Goldstein:2017bua,german3}. Then, the total von Neumann entropy or EE can be written as a sum of the contributions from these different symmetry sectors \cite{Goldstein:2017bua,german3}. As each of these blocks can individually be diagonalized, then from the eigenvalues coming from each block, one can define a notion of EE associated with a definite symmetry sector. It was termed Symmetry-Resolved Entanglement Entropy (SREE). This notion turns out to be particularly useful for studying the role of internal symmetry in quantum field theories and quantum many-body systems, and there are possibilities that it can be experimentally measured \cite{Islam,Goldstein:2017bua,expSRE1,expSRE2, expSRE3, expSRE4}, providing further motivation to study this quantity. This has also been extended to Renyi entropies, which are termed Symmetry-Resolved Renyi Entropies (SRRE) \cite{Goldstein:2017bua}.
\medskip

The idea behind the computation of SRRE for a particular CFT with internal symmetry $U(1)$ or $\mathbb{Z}_k$ (with $k$  being an integer)  is simple: One needs to compute the partition function on the replicated manifold in the presence of an Aharonov-Bohm flux. Finally, it boils down to the calculation of the charge moments for the subsystem with the help of composite twist operators \cite{Goldstein:2017bua}. In \cite{Belin:2013uta,Belin:2014mva,Matsuura:2016qqu}, such computations were first presented (albeit using a different normalization), and the physical meaning of SREE is further clarified in \cite{Goldstein:2017bua}. Subsequently, SREE has been studied for ground states of various models of CFTs, free massive QFTs, critical spin chains, and certain integrable spin chains, for excited states of certain CFTs and free fermionic and bosonic models, as well as for certain quench dynamics \cite{Goldstein:2017bua,german3,Fraenkel:2019ykl,Bonsignori:2020laa,Capizzi_2020,Estienne:2020txv,Ares:2022gjb,Fossati:2023zyz,Gaur:2023yru,Ares:2022koq,Ares:2022hdh,Horvath:2020vzs,Horvath:2021fks,CTFMichele,freeboson,ourPartI,Horvath:2021rjd,Murciano:2019wdl,FG,Murciano:2020lqq,Parez:2021pgq,Fraenkel:2021ijv,Ghasemi:2022jxg,Baiguera:2022sao,di2023boundary,Kusuki:2023bsp,Pirmoradian:2023uvt,Gaur:2024vdh}. These computations have been extended for the Wess-Zumino-Witten model, as well as for theories with non-invertible symmetries \cite{Calabrese:2021wvi,Sierra_2024}. Furthermore, SREE has been studied for disordered systems, many-body localized systems, and for certain topological phases \cite{chen2023energy, Turkeshi:2020yxd, Kiefer_Emmanouilidis_2020,Monkman:2020ycn,Azses:2020wfx,Cornfeld:2018wtp,Ara:2023pnn}. Not just in computing SREE, but the idea of symmetry resolution has also been extended to exploration of other measures of entanglement, e.g., Computable Cross-Norm (CCNR) negativity, relative entropies, and fidelities \cite{Murciano:2021djk,Chen:2021pls,Capizzi:2021zga,Parez:2022sgc,ourPartIII,Berthiere:2023gkx,Bruno:2023tez}. Last but not least, symmetry-resolved entanglement measures have found intriguing applications in the context of holography (focusing on the AdS/CFT correspondence)\cite{Belin:2013uta,Belin:2014mva, Zhao:2020qmn,Zhao:2022wnp, Weisenberger:2021eby} \footnote{The SREE references in this paragraph are not exhaustive. For more references and details, interested readers are referred to the review \cite{Castro-Alvaredo:2024azg}.}. 
\medskip

Despite all this, exact analytical computations  for SREE are not too well explored as of yet, except for certain CFTs and QFTs. For example, CFT in $1+1$d with global $U(1)$ charge symmetries could have infinite dimensional symmetries courtesy of the Virasoro-Kac-Moody algebra. The motivation of the current work starts with adding similar computations for Non-Lorentzian conformal field theories in 2d, namely Galilean and Carrollian conformal field theories (GCFTs and CCFTs), with added global charges to the arsenal. These theories arise from Inönü-Wigner contractions of CFTs \cite{Bagchi:2009pe} in the same dimensions and have been widely used to understand various physical situations, including the holographic principle in the flat-space setting. GCFTs and CCFTs appear as non-relativistic and ultra-relativistic limits of known conformally invariant theories \cite{Bagchi:2009my, Bagchi:2009pe,Duval:2014uoa}, but could also be regarded as intrinsically defined theories with Galilean/Carrollian symmetries. However, in 1+1 dimensions they are exactly isomorphic under the exchange of space and time directions. In fact, both of these theories are governed by the same symmetry algebra,
\begin{align}\label{BMS3q}
[L_n,L_m]=&(n-m)L_{n+m} + \frac{c_L}{12}(n)(n^2-1)\delta_{n+m}, \nonumber\\
[L_n,M_m]=&(n-m)M_{n+m} +\frac{c_M}{12}(n)(n^2-1)\delta_{n+m}, \\
[M_n,M_m]=&0 \nonumber.
\end{align}
where $c_L$ and $c_M$ are central charges. The generators $L,M$ come from two different contractions of 2d Virasoro generators (see Appendix~(\ref{AppB})) and bear different physical meanings. In this work, we will be looking at SREE calculations in these theories, and also for the same with added global $U(1)$ symmetries in the boundary theory, i.e. for Non-Lorentzian Kac-Moody algebras \cite{Bagchi:2023dzx}. 
\medskip

Another intriguing motivation to study SREE in non-Lorentzian theories comes from the holography of asymptotically flat spacetimes, as we have already mentioned. The idea in a nutshell is: $d$-dimensional Conformal Carrollian Algebras (CCAs) are isomorphic to asymptotic symmetry algebras of $(d + 1)$ dimensional flat spacetimes, known as the BMS algebra \cite{Bondi:1962px, Sachs:1962zza}. CCFTs (and GCFTs in 2d) can thus act as putative dual theories of flat spacetimes  \cite{flatHol,flatHol1,flatHol2,flatHol3,flatHol4,flatHol7,Bhattacharyya:2023czi}. 
Not only that, very recent understanding of maps between 3d Carroll CFTs and scattering amplitudes in 4d asymptotically flat spacetimes \cite{Bagchi:2022emh, Donnay:2022wvx, Bagchi:2023cen} has opened up new horizons and connections to the study of so-called celestial holography (see \cite{Raclariu:2021zjz, Pasterski:2021rjz} for recent reviews), the flat space theory for codimension-two holography. In this connection, EE for GCFTs and CCFTs from both sides of the duality has been explored in detail \cite{Basu:2015evh,Basu:2017aqn,Grumiller:2019xna,Jiang:2017ecm} over the years. It then makes perfect sense to explore the new measure of entanglement in the block, i.e. SREE/SRRE for such non-Lorentzian theories, and to try working out the putative interpretation in Flat holography. Since CFT$_2$ shows universal results as far as SREE/SRRE is concerned in the sense of \textit{equipartition of entanglement} among all charge sectors, we would also be keeping an eye out for such insights from our investigation.

\medskip

The organization of the paper is as follows: In Sec.~(\ref{sec2}), we briefly review the essential
facts about SREE and SRRE and how to compute them using the replica trick for a CFT$_2$. In Sec.~(\ref{sec3}), we extend the computation of SREE and SRRE for a boosted interval in $1 + 1$d CFT.
In Sec.~(\ref{sec4}), we consider the non-relativistic limit of the result for the boosted interval to
obtain the SREE result for GCFT. In Sec.~(\ref{sec5}), we provide an intrinsic computation of SRRE
and SREE for GCFT and provide a match with the result coming from the limiting procedure. Then in the Sec.~(\ref{sec6}), we consider the ultra-relativistic limit of the CFT results
to obtain the SEE for BMS$_3$(Carrollian) CFT which is relevant for flat space holography. In Sec.~(\ref{sec7}), we extend the computation for BMS$_3$-Kac-Moody theory and comment on the connection with flat holography. Finally, we summarize our results and provide future outlooks in Sec.~(\ref{sec8}). Some details helpful for understanding the setup and some of the computation not presented explicitly in the main text are provided in the Appendix~(\ref{AppB}) and (\ref{AppA}).

\section{Replica method and symmetry resolution} \label{sec2}
The von Neumann entanglement entropy, hereafter referred to as EE, of a subregion $A$ for a many degrees of freedom system, is specific to a state represented by a density matrix $\hat{\rho}$. The first step is to define the reduced-density matrix $\hat{\rho}_A = \mathrm{Tr}_B \hat{\rho}$ for the degrees of freedom residing in the subregion $A$, \textcolor{black}{by tracing out the degrees of freedom 
of the complementary region $B$}. Then
\textcolor{black}{\begin{equation}\label{Renyii} S(A)=\lim_{n\to 1} \frac{1}{1-n}\ln\textrm{Tr}(\hat{\rho}^{n}_{A})\end{equation}}
gives the desired EE as described above. 
\medskip

For 2d CFTs, the above computation is accomplished by \cite{Calabrese:2004eu, Calabrese:2009qy} a very useful approach well-known as the replica trick. One defines the CFT on an $n-$sheeted copy of the original Riemann surface, glued at an equal time slice of the region $A$. If $Z_n(A)$ is the partition function on the replicated manifold and $Z$ is that of the original one, then as per \cite{Calabrese:2009qy}  
\begin{equation} 
\textrm{Tr}(\hat{\rho}_{A}^{n})= Z_{n}(A)/Z^{n}\,, \label{EE}\end{equation}
where $Z_{n}(A)\propto \langle \mathcal{T}_{n}(u,0) \tilde{\mathcal{T}}_{n}(v,0)\rangle. $
Here the $\mathcal{T}_{n}, \tilde{\mathcal{T}}_{n} $  are the twist operators inserted at the end of the subregion with the scaling dimension $ d_{n}=\frac{c}{12}(n-\frac{1}{n}).$ In this paper, we will mainly consider the subregion as a single interval  and restrict ourselves to $ 1+1$d CFT at \textcolor{black}{zero temperature}. Now it follows from \cite{Calabrese:2009qy} that, 
\begin{equation}\label{t=0}  \textrm{Tr}(\hat{\rho}_{A}^{n})=c_n \Big|\frac{v-u}{\varepsilon}\Big|^{-\frac{c+\bar{c}}{12}(n-\frac{1}{n})}\,, \end{equation}
where, $c_{n}$ is a constant to make it dimensionless with $c_1=1$ and $\varepsilon$ is the UV-cutoff for the theory. Also, as usual, $c$ and $\bar{c}$ are the two central charges coming from the two copies of the Virasoro algebra, and we will assume throughout this paper that they are not necessarily equal to each other. Then from (\ref{EE}) it follows that the EE of the subregion can be written as:

\textcolor{black}{\begin{equation}
S(A)=\frac{c+\bar{c}}{6}\ln\Big(\frac{L}{\varepsilon}\Big)+\mathfrak{b}\,,
\end{equation}}

where, $L=|u-v|$ is the size of the interval (we are working with the $t=0$ slice in particular) and $\mathfrak{b}$ denotes non-universal contributions to the EE. We will focus only on the universal contribution to $S(A)$ in what follows. 
\medskip

\par
\textcolor{black}{Now, let us turn our attention to the specific computation we are concerned about in this work, i.e. that of $\textrm{Tr}(\hat{\rho}_{A}^{n})$ when an Aharonov-Bohm flux is inserted in the replicated manifold between the Riemann sheets. Let $\hat{N}$ be the total conserved charge for a particular global symmetry. For local theories and symmetry generators having local transformation properties, it is reasonable to assume the existence of the conserved charges supported on subsystems $A$ and $B$ individually, such that ${\hat{N}}={\hat{N}}_A+{\hat{N}}_B$, i.e., the sum of the contribution to the subsystem $A$ and its complement ($B$).
We are interested in a state given by the density matrix $\hat{\rho}$ of $A \cup B$, which has a definite charge with respect to $\hat{N}$, i.e. $[{\hat{N}}, \hat{\rho}] = 0$. Tracing this over the Hilbert space supported in the subregion $B$, we get $[{\hat{N}}_A, \hat{\rho}_A] = 0$. It then immediately follows that the reduced density matrix for subsystem $A$, $\hat{\rho}_A$ takes a block diagonal structure where each block corresponds to each of the charge sectors (for a particular value of $N_A$) for the subregion $A$, i.e. $\hat{\rho}_A = \bigoplus_{N_A} \hat{\Pi}_{N_A}\hat{\rho}_A,$ where $\hat{\Pi}_{N_A}$ is a operator which projects  $\hat{\rho}_A$ to a particular charge sector corresponding to a fixed value of $N_A$ \cite{Bonsignori:2019naz}.\footnote{$N_A$ denotes the eigenvalues of the operator $\hat{N}_A.$} 
Now the symmetry resolved Renyi/Entanglement entropies (SRRE/SREE) can be defined as \cite{Goldstein:2017bua,german3, Bonsignori:2019naz}, 
\begin{equation} \label{fluxpartition2}
    S_n =  \frac{1}{1-n}\ln\left[\frac{\mathcal{Z}_n(N_A)}{\mathcal{Z}_1(N_A)^n} \right];~~~~S(N_A) = -\partial_n\left[\frac{\mathcal{Z}_n(N_A)}{\mathcal{Z}_1(N_A)^n} \right]_{n\to 1}
\end{equation}
where the charged moments of the density matrix $\mathcal{Z}_n(N_A)$ are defined as \cite{Goldstein:2017bua,german3}, 
\begin{align} \label{fluxpartition}
\mathcal{Z}_{n}(N_A)=\textrm{Tr}(\hat{\Pi}_{N_A}\hat{\rho}^n_{A})=\int_{-\pi}^{\pi} \frac{d\alpha}{2\pi} e^{-i\alpha\, N_A} Z_n (\alpha)\,, \quad Z_{n}(\alpha)=\textrm{Tr}({\hat{\rho}}^n_A e^{i\,\alpha\,{\hat{N}}_A})\,.
\end{align}
$Z_n(\alpha)$ has an interpretation as a partition function on the $n$-sheeted Riemann surface in the presence of the Aharonov-Bohm flux $\alpha$, coupled to the charge corresponding to the particle number ($N_A$) in the subsystem $A$. Note that here $Z_n(0) =\textrm{Tr}({\hat{\rho}}^n_A) $, and by calculating $Z_n(\alpha)$ for a system, we can easily compute the SRRE and SREE associated with it \cite{Goldstein:2017bua}. In the remainder of the paper, we will mainly consider $U(1)$ charges coupled to our theory of interest, which will make the above computation more tractable.}
\medskip

\textcolor{black}
{Moreover, we would also be interested to understand how the entanglement is distributed within the different charge sectors, for which one can decompose the von Neumann entropy ($ S_{\textrm{vN}}$) in the following way \cite{Bonsignori:2019naz}: 
\begin{equation} \label{fluxpartition1}
    S_{\textrm{vN}}= S^c +S^f = \sum_{N_A}S(N_A)P_1(N_A) - \sum_{N_A}P_1(N_A)\text{ln} P_1(N_A)
\end{equation}
The two different contributions in \eqref{fluxpartition1} are respectively called the \textit{configurational} entanglement entropy $S^c$, which measures the entropy of each charge sector, and \textit{fluctuation} entanglement entropy $S^f$, which measures the fluctuations of the value of the charge (or number of particles) within the subsystem. Also, note that $Z_1(\alpha)$ in \eqref{fluxpartition} is the generating function of the charge distribution and $Z_1(N_A)$ can be easily read off as the probability $P_1(N_A)$ of having $N_A$ number of particles in the subregion $A$ \cite{Goldstein:2017bua,Bonsignori:2019naz}. This decomposition will make sure a non-trivial cancellation happens between these two quantities that only leads to the von Neumann terms.  }
\medskip

Now, let us quickly review the way to compute $Z_{n}(\alpha)$ for a CFT  ground state using the replica trick. The key point here is that, in the presence of flux, the local twist field $\mathcal{T}$, used to generate the twisted boundary conditions, must be fused with the operator generating the flux $\mathcal{V}$. Moreover, a composite twist field $\mathcal{T}_{\mathcal{\nu}}=\mathcal{T}\mathcal{V}$ is needed in this case instead of just $\mathcal{T}$, the latter of which is sufficient only in the absence of this additional flux. We can show that this composite twist field behaves as a primary operator with scaling dimension $\Delta_n$, which can then be extracted from a one-point function of the stress-energy tensor on this $n$-sheeted Riemann surface with the Aharonov-Bohm flux ($\mathcal{R}_{n,\alpha}$). Explicitly, we can then write \cite{Goldstein:2017bua}:
\begin{align}
    \begin{split}
\langle{T}(z)\rangle_{\mathcal{R}_{n,\alpha}} =\frac{\langle {T}(z) \mathcal{T}_{\nu}(w) \mathcal{\tilde{T}}_{\nu}(w^{\prime})\rangle_{\mathcal{C}}} {\langle \mathcal{T}_{\nu}(w) \mathcal{\tilde{T}}_{\nu}(w^{\prime})\rangle_{\mathcal{C}}} 
=\Delta_{n}\frac{(w-w^{\prime})^{2}}{(z-w)^{2}(z-w^{\prime})^{2}}\,,\label{twist}
\end{split}
\end{align}
where $\Delta_{n}$ is the scaling dimension and it is given by,
\begin{equation} \Delta_{n}= \frac{c}{24}\left(n-\frac{1}{n}\right)+\frac{\Delta_{\mathcal{V}}}{n}\,.\end{equation}
 Here, one has to use a uniformization map 
 to go from $\mathcal{R}_{n,\alpha}$ to the complex plane $\mathcal{C}$, where we insert the operators at different coordinates $z,w,w'$. $\Delta_{\mathcal{V}}$ is the scaling dimension of the field that generates the flux, and it is, in general, a function of the flux parameter $\alpha.$ For $1+1$d CFT, this is typically that of a vertex operator, which takes the form $\frac{1}{2}(\frac{\alpha}{2\pi})^2\,K\,.$ The crucial point is that the composite twist operator $\mathcal{T}_{\mathcal{\nu}}$ also behaves as a primary operator. 
 \medskip

 To determine the precise form of $\Delta_{\mathcal{V}}$ (i.e. $K$), one has to consider some particular models. Interested readers are referred to \cite{german3,Goldstein:2017bua, Bonsignori:2019naz,Capizzi_2020,Calabrese:2021wvi, Castro-Alvaredo:2024azg} for various theory-specific studies. Finally, we can use the two-point correlator of the composite twist field $\mathcal{T}_{\mathcal{\nu}}$ to obtain $Z_n(\alpha)$ as defined in (\ref{fluxpartition}). For $1+1$d CFT we get:
 
\textcolor{black}{\begin{equation}
    Z_{n}(\alpha) \sim L^{-\frac{c+\bar{c}}{12}(n-\frac{1}{n})}L^{-\frac{2}{n}(\Delta_{\mathcal{V}}+\bar{\Delta}_{\mathcal{V}})}\,,
\end{equation}}
Here, we consider contributions from both holomorphic and anti-holomorphic sectors. The $\bar{\Delta}_{\mathcal{V}}$ is the scaling dimension of the anti-holomorphic part of $\mathcal{V}$. As mentioned in (\ref{fluxpartition}), we can now obtain the charged moment $\mathcal{Z}_n(N_A)$ after performing the Fourier transform. Then using (\ref{fluxpartition2}) one can get $S_n(N_A)$ . This gives us the way to calculate symmetry resolved Renyi entropy (SRRE) once $Z_{n}(\alpha)$ is known. 
\medskip

Given this setup, we will next study this SRRE as well as the SREE for a boosted interval in $1+1$d CFT, which will be the defining calculation for this work. In what follows next, we will use those results to investigate various extreme limits of the boost parameter, thereby generating these results to their Non-Lorentzian cousins. 
\section{Symmetry resolution for CFT with boosted interval}\label{sec3}


In this section, we first briefly discuss the structure of the RE and EE of the ground state for a boosted interval in a $1+1$d CFT following \cite{Wall:2011kb,Castro:2014tta}. We will denote the endpoints of the interval as $1$ and $2$, respectively. To calculate the RE, we need to compute the partition function $Z_n(A)$ as given in (\ref{EE}). This is determined by the two-point correlators of the twist field, as discussed earlier. One gets by combining holomorphic and anti-holomorphic contributions: 
\begin{equation} \label{c}
   Z_{n}=\Big(\frac{z_{12}}{\varepsilon}\Big)^{-\frac{c}{12}(n-\frac{1}{n})} 
   \Big(\frac{\bar{z}_{12}}{\varepsilon}\Big)^{-\frac{\bar{c}}{12}(n-\frac{1}{n})}
\end{equation}
where $z_{12}=z_1-z_2,\bar{z}_{12}=\bar{z}_1-\bar{z}_2\in \mathcal{C}$ and they are complex conjugate of each other. As introduced earlier, $\varepsilon$ is the UV cutoff. If there is no boost, we can choose a constant time slice and in that case, $|z_{12}|=|\bar{z}_{12}|=L\,,$ where $L$ is the spatial separation of the endpoints of the interval, and then we immediately get back (\ref{t=0}).
\medskip

Now, the RE for the single interval is given by
\begin{equation} \label{Renyi}
S_{n}=\frac{\textrm{ln}\Big(\textrm{Tr}\rho^n_A\Big)}{1-n}=\Big(1+\frac{1}{n}\Big)\Big[\frac{c}{12}\ln\Big(\frac{z_{12}}{\varepsilon}\Big)+\frac{\bar{c}}{12}\ln\Big(\frac{\bar{z}_{12}}{\varepsilon}\Big)\Big]\,.\end{equation}
Then, the EE can be written as:
\begin{equation} \label{boostEE} S(A)=\lim_{n\rightarrow 1}S_n= \frac{c+\bar{c}}{6}\ln\Big(\frac{L}{\varepsilon}\Big)+i\,\Big(\frac{c-\bar{c}}{6}\Big)\,\theta, \end{equation}
where we have parametrized the interval $z_{12}$ as $z_{12}=L\,e^{i\,\theta}\,.$ Note that until now, all computations have been done using Euclidean correlators. We now analytically continue the final expression to the real-time value, which amounts to mapping $\theta$ to the proper boost parameter $\kappa$ as
$\theta=-i\kappa\,.$
The the result (\ref{boostEE}) transforms to, 
\begin{equation} S(A)=\frac{c+\bar{c}}{6}\ln\Big(\frac{L}{\varepsilon}\Big)+\,\Big(\frac{c-\bar{c}}{6}\Big)\,\kappa\,.
\end{equation}
Notice that, due to the presence of a non-vanishing boost, we get a contribution proportional to the difference between the central charges ($c-\bar{c}$) in the EE. This will be crucial when we discuss symmetry resolution for non-relativistic and ultra-relativistic theories, where this situation occurs naturally. 
\medskip

Before we discuss these two limits, it is instructive to write the partition function in the presence of the Aharonov-Bohm flux $\alpha,$ which generalises from \eqref{c}: 

\begin{equation} \label{sboost}
\textcolor{black}{Z_n}(\alpha)=\Big(\frac{z_{12}}{\varepsilon}\Big)^{-\frac{c}{12}(n-\frac{1}{n})-\frac{2\,\Delta_{\mathcal{V}}}{n}}\Big(\frac{\bar{z}_{12}}{\varepsilon}\Big)^{-\frac{\bar{c}}{12}(n-\frac{1}{n})-\frac{2\,\bar{\Delta}_{\mathcal{V}}}{n}}\,,
\end{equation}
where $\Delta_{\mathcal{V}}$ and $\bar{\Delta}_{\mathcal{V}}$ are the scaling dimension of the holomorphic and anti-holomorphic part of the composite twist operator as defined in (\ref{twist}). After analytically continuing to Lorenztian time, we can again parameterize the interval in terms of the boost: $z_{12}=L e^{\kappa}= L\,\cosh(\kappa)+L\,\sinh(\kappa).$ Then it is easy to make the following identifications: 
$$\beta=\frac{v}{c}=\tanh(\kappa),\quad \gamma=\frac{1}{\sqrt{1-\beta^2}}=\cosh(\kappa),\quad \gamma\,t_{12}=L\, \cosh(\kappa),\quad \gamma \beta=L\,\sinh(\kappa)\,.$$

Here, $x_{12}$ and $t_{12}$ denote the spatial and temporal separation of the endpoints of the boosted interval (subregion $A$). Also, after the analytic continuation to Lorentzian time, we can write the boosted relations $z_{12}=\gamma(t_{12}+\beta x_{12})$ and $\bar{z}_{12}=\gamma(t_{12}-\beta x_{12})\,.$ Using these we can then rewrite (\ref{sboost}) in the following way, 
\begin{align}
\begin{split}
\textcolor{black}{Z_n}(\alpha)&=\Big[\frac{\gamma}{\varepsilon}(t_{12}+\beta x_{12})\Big]^{-\frac{c}{12}(n-\frac{1}{n})-\frac{2\,\Delta_{\mathcal{V}}}{n}}\Big[\frac{\gamma}{\varepsilon}(t_{12}-\beta x_{12})\Big]^{-\frac{\bar{c}}{12}(n-\frac{1}{n})-\frac{2\,\bar{\Delta}_{\mathcal{V}}}{n}}\,,\\&
=\Big(\frac{\gamma}{\varepsilon}\Big)^{-\Big[\frac{c+\bar{c}}{12}(n-\frac{1}{n})+\frac{2\,(\Delta_{\mathcal{V}}+\bar{\Delta}_{\mathcal{V}})}{n}\Big]}(t_{12})^{-\Big[\frac{c+\bar{c}}{12}(n-\frac{1}{n})+\frac{2\,(\Delta_{\mathcal{V}}+\bar{\Delta}_{\mathcal{V}})}{n}\Big]}
\\&\hspace{0.5cm}\Big(1+\beta\frac{x_{12}}{t_{12}}\Big)^{-\frac{c}{12}(n-\frac{1}{n})-\frac{2\,\Delta_{\mathcal{V}}}{n}}\Big(1-\beta\frac{x_{12}}{t_{12}}\Big)^{-\frac{\bar{c}}{12}(n-\frac{1}{n})-\frac{2\,\bar{\Delta}_{\mathcal{V}}}{n}}\,.\label{limit}
\end{split}
\end{align}
The partition function in the form \eqref{limit} is ideal for discussing the non-relativistic/ultra-relativistic limits.

\section{From CFT to GCFT: Symmetry resolution} \label{sec4}
In this section, starting from the techniques that have been built as of now, we will investigate the symmetry resolution for the ground state of a GCFT$_2$. We will focus particularly on computing the symmetry-resolved entropy for our investigations. First, we will do it by taking the non-relativistic limit of the CFT result mentioned in (\ref{limit}). Since we now have the results for the boosted intervals, our approach will be slightly different from that of \cite{Bagchi:2014iea}, in the sense that we will be using $\beta$ as an effective contraction parameter, instead of inhomogeneously scaling space and time.
As we know, in the Galilean limit \cite{Bagchi:2009pe} the speed of light goes to infinity, i.e. for the Lorentz parameters,
$$ \beta=\frac{v}{c}\rightarrow 0 ,~~\gamma=\frac{1}{\sqrt{1-\beta^{2}}}\rightarrow 1\,. $$
This will be the explicit limit we will be taking on our CFT results.
Also, as argued in \cite{Bagchi:2009pe}, we have to consider the central charges $c$ and $\bar{c}$ to be very large. 
Then, using the fact that $\lim_{m\to\infty}(1+x/m)^{-m}=e^{-x},$
we can rewrite (\ref{limit}) as follows:
\begin{align}
    \begin{split} \label{compare1}
\textcolor{black}{Z_n}(\alpha)=\Big(\frac{t_{12}}{\varepsilon}\Big)^{-\left[\frac{c+\bar{c}}{12}(n-\frac{1}{n})+\frac{2\,(\Delta_{\mathcal{V}}+\bar{\Delta}_{\mathcal{V}})}{n}\right]} \exp\Big[-\Big\{\frac{\beta(c-\bar{c})}{12}\Big(n-\frac{1}{n}\Big)+\frac{2\,\beta\,(\Delta_{\mathcal{V}}-\bar{\Delta}_{\mathcal{V}})}{n}\Big\}\frac{x_{12}}{t_{12}}\Big]\,.
\end{split}
\end{align}
At this point, for simplicity we can set the UV-cutoff $\varepsilon\rightarrow 1 $ and identify the Galilean parameters under the limit $\beta \to 0$ \cite{Bagchi:2009pe,Bagchi:2014iea,Basu:2015evh}:
\begin{equation} \label{comapre1}
c+\bar{c}=c_{L},\quad \beta(c-\bar{c})=c_{M},\quad \Delta_{\mathcal{V}_{1}}=\Delta_{\mathcal{V}}+\bar{\Delta}_{\mathcal{V}},\quad \Delta_{\mathcal{V}_{2}}=\beta(\Delta_{\mathcal{V}}-\bar{\Delta}_{\mathcal{V}})\,.
\end{equation}
Note that, as mentioned earlier, the boost parameter $\beta$ itself becomes the contraction parameter for our case. In fact, it is easy to see that the limit $\beta\to 0,~\gamma \to 1$ implies the well-known Galilean contraction $t\to t,~x\to \epsilon x,~\epsilon \to 0$. One should compare these limits with Appendix \eqref{AppB} for more clarity.
Combining everything, we get a $\beta$ independent result for symmetry resolved moments in GCFT:
\begin{equation} \label{sgcft}
\textcolor{black}{Z_n}(\alpha)= {t_{12}}^{-\left[\frac{c_L}{12}(n-\frac{1}{n})+\frac{2\,\Delta_{\mathcal{V}_{1}}}{n}\right]} \exp\Big[-\Big\{\frac{c_M}{12}\Big(n-\frac{1}{n}\Big)+\frac{2\,\Delta_{\mathcal{V}_{2}}}{n}\Big\}\frac{x_{12}}{t_{12}}\Big]\,.
\end{equation}
Notice that we have already assumed that $c, \bar{c} \rightarrow \infty$ and $\beta \rightarrow 0$ such that $\beta(c-\bar{c})$ is finite in this limit. 
\medskip

Now we will calculate the full SREE for GCFT. Typically, to incorporate the effect of the accumulated phase due to the flux ($\alpha$) in the path integral, and subsequently in the SREE, we need to insert the following vertex operators at the end points of the interval on the $j$th copy of the replicated manifold:
$$ \mathcal{V}_{1}=e^{i\frac{\alpha}{2\pi}\phi_{j}} ,\mathcal{V}_{2}=e^{i\frac{\alpha}{2\pi}\phi_{j^{\prime}}}\,.$$
Consequently, we get the dimensions of the composite twist operators:
\begin{equation} \label{vert}
\Delta_{\mathcal{V}_{1}}=\frac{1}{2}\Big(\frac{\alpha}{2\pi}\Big)^{2} K_{1},~~\Delta_{\mathcal{V}_{2}}=\frac{1}{2}\Big(\frac{\alpha}{2\pi}\Big)^{2} K_{2}\,.
\end{equation}
Note that the constants $K_1$ and $K_2$ depend on the underlying theory and we will consider certain examples later on while taking various limits of this CFT result.
\footnote{As an example, if we consider the Luttinger liquid described by a $c=1$ CFT, then $\Delta_{\mathcal{V}}=\bar{\Delta}_{\mathcal{V}}=\frac{1}{2}(\frac{\alpha}{2\pi})^{2} K,$ where $K$ is the Luttinger liquid parameter. Then it readily follows: $K_1=2 K, K_2=0\,.$}. Using this fact, we can rewrite (\ref{sgcft}) as follows: 
\begin{equation}
\textcolor{black}{Z_n}(\alpha)=t_{12}^{-\left[(n-\frac{1}{n})\frac{c_{L}}{12}+(\frac{\alpha}{2\pi})^{2}\frac{K_{1}}{n}\right]} \exp\Big[-\Big\{\Big(n-\frac{1}{n}\Big)\frac{c_{M}}{12}+\Big(\frac{\alpha}{2\pi}\Big)^{2}\frac{K_{2}}{n}\Big\}\frac{x_{12}}{t_{12}}\Big]\,.
\end{equation}
We further denote 
\begin{equation}
  K_{1}\ln t_{12}+K_{2}\frac{x_{12}}{t_{12}}= L' \,.  
\end{equation}

Then the charged moment $\textcolor{black}{\mathcal{Z}_n}(N_{A})$ can now be written as the Fourier transform of $\textcolor{black}{Z_n}(\alpha)$ as shown in (\ref{fluxpartition}). We will work in the limit $L'\gg 1\,$ to write:
\begin{align}
\begin{split} \label{PGCFT}
\textcolor{black}{\mathcal{Z}_n}(N_{A})&\approx\frac{\textcolor{black}{Z}_{n}(\alpha=0)}{2\pi}\int_{-\infty}^{\infty}t_{12}^{-(\frac{\alpha}{2\pi})^{2}\frac{K_{1}}{n}} \exp\left(-\left(\frac{\alpha}{2\pi}\right)^{2}\frac{K_{2}\, x_{12}}{n\,t_{12}}\right) e^{-i\alpha N_{A}}\,d\alpha\\&
=\frac{Z_{n}(\alpha=0)}{2\pi}\int_{-\infty}^{\infty}e^{-\frac{\alpha^{2}\,L'}{4\pi^{2}n}} e^{-i\alpha N_{A}}\,d\alpha\\&
=\frac{Z_{n}(\alpha=0)}{2\pi}\sqrt{\frac{4\pi^{3}\,n}{L'}} e^{-\frac{\pi^{2}N_{A}^{2}n}{L'}}\,,
\end{split}
\end{align}
where the zero flux moment is given by:
\begin{equation}
\textcolor{black}{Z}_{n}(\alpha=0)=t_{12}^{-\frac{c_{L}}{12}(n-\frac{1}{n})} e^{-\frac{c_{M}}{12}(n-\frac{1}{n})\frac{x_{12}}{t_{12}}}\,
\end{equation}

Note that the range of the original integration as mentioned in (\ref{fluxpartition}) was from $-\pi$ to $\pi.$ But as we are working in large $L'$ limit, we can show that the integral mentioned in (\ref{PGCFT}) 
can be well approximated by the saddle-point value and will decay fast outside this interval\footnote{In general the integral between $-\pi$ to $\pi$ gives rise to an error function ($\erf[L']$)\cite{Bonsignori:2019naz,Castro-Alvaredo:2024azg}\,.}. Hence, we can extend the range of the integral covering the entire real line and perform the Gaussian integral. 
We can now calculate the SREE using the expressions in (\ref{fluxpartition2}),
\begin{align}
\begin{split} \label{SREEGCFT}
S_{\textrm{GCFT}}(N_{A})=\frac{c_{L}}{6}\ln{t_{12}}+\frac{c_{M}}{6}\frac{x_{12}}{t_{12}}-\frac{1}{2}+\frac{1}{2} \ln \left(\frac{\pi}{L'}\right)\,.
\end{split}
\end{align}
Notice that, we have kept the $\mathcal{O}(1)$ and the subleading (in $L'$) term as well. As discussed in Sec.~(\ref{sec2}), $S_{1}(N_{A})=P_1(N_{A})$, where $P_1(N_{A})$ is the probability of having $N_{A}$ number of particles in region $A$. For our case, from (\ref{PGCFT}) we have the following, 
\begin{align}
\begin{split}
P_1(N_{A})=\sqrt{\frac{\pi}{L'}}e^{-\frac{\pi^{2}N_{A}^{2}}{L'}}\,.\end{split}
\end{align}
\textcolor{black}{Now to get the total EE we make use of (\ref{fluxpartition1}). 
The configurational entropy comes out to be,
\begin{equation}
    S_{\textrm{GCFT}}^{c}=\int_{-\infty}^{\infty}S_{\textrm{GCFT}}(N_{A}) P_{1}(N_{A})\,dN_{A}=\frac{c_{L}}{6}\ln{t_{12}}+\frac{c_{M}}{6}\frac{x_{12}}{t_{12}}-\frac{1}{2}+\frac{1}{2} \ln \left(\frac{\pi }{L'}\right)\,.
\end{equation}
Also the fluctuation entropy can be computed as,
\begin{equation}
    S_{\textrm{GCFT}}^{f}=-\int_{-\infty}^{\infty}P_{1}(N_{A})\ln{P_{1}(N_{A})}\,dN_{A}= \frac{1}{2}+\frac{1}{2} \ln \left(\frac{L' }{\pi}\right)\,.
\end{equation}
Then we can write the total EE using (\ref{fluxpartition1}),
\begin{equation} \label{TotalEntropyGCFT}
   S_{\textrm{GCFT}}(A)=S_{\textrm{GCFT}}^{c}+S_{\textrm{GCFT}}^{f}=\frac{c_{L}}{6}\ln{t_{12}}+\frac{c_{M}}{6}\frac{x_{12}}{t_{12}}\,.
\end{equation}}

This matches exactly the previous result for the EE for GCFT as derived in \cite{Bagchi:2014iea,Basu:2015evh}. Note that the contributions from the $\mathcal{O}(1)$ term and the subleading term, outright cancel  and only the first two terms as in (\ref{TotalEntropyGCFT}) remain. Therefore, (\ref{TotalEntropyGCFT}) serves as a consistency check of our computation.

\section{Symmetry resolution for GCFT: An intrinsic computation}\label{sec5}

 Having obtained the result for the SREE for the ground state of the GCFT by taking the appropriate limit of the 2d CFT result, we now set up the same computation from an intrinsic viewpoint using the GCFT$_2$ correlators.  The intrinsic symmetries in a Galilean $1+1$d CFT can be encoded into two energy-momentum tensors \cite{Bagchi:2010vw},
\begin{align}
\begin{split}   \label{T12}
&T_{1}(x,t)=\sum t^{-n-2}\Big[L_{n}+(n+2)\frac{x}{t}M_{n}\Big]\,,\\&
T_{2}(x,t)=\sum t^{-n-2}M_{n}
\end{split}
\end{align}
where $L_{n}$ and $M_{n}$ are generators of the $1+1$d GCA \cite{Bagchi:2009pe} as mentioned in \eqref{BMS3q}. The primary fields are characterized by two distinct weight factors: $\{h_{L},h_{M}\}\,,$ which are determined by the Ward identities of the primary operators with $T_1$ and $T_2$ respectively. As discussed in Sec.~(\ref{sec2}), to compute the symmetry-resolved entropy, we will need a two-point correlator for the composite twist operators. Note that these twist operators are primary fields. For GCFT, the two-point correlators of the primary fields $\phi_1(x_1,t_1)$ and $\phi_2(x_2,t_2)$ are given by \cite{Bagchi:2009pe}, 
\begin{equation} \label{GCFT2pt}
\langle \phi_{1}\phi_{2}\rangle=t_{12}^{-2h_{L}} \exp\Big(-2h_{M}\frac{x_{12}}{t_{12}}\Big)\,.
\end{equation}
For the composite twist operators ($\mathcal{T}_{\mathcal{\nu}_1}=\mathcal{V}_1\mathcal{T}$ and $\mathcal{\bar{T}}_{\mathcal{\nu}_2}=\mathcal{V}_2\mathcal{\bar{T}}$) \textcolor{black}{inserted on the complex plane} 
in presence of the Aharonov-Bohm flux, $h_L$ and $h_M$ can be obtained from the one-point function of $\langle{T_1}\rangle_{\mathcal{R}_{n,\alpha}}$ and $\langle{T_2}\rangle_{\mathcal{R}_{n,\alpha}}$ as discussed in Sec.~(\ref{sec2}). They turn out to be the following: 
\begin{equation}\label{Appendix}
h_L= \Big(n-\frac{1}{n}\Big)\frac{c_{L}}{24}+\frac{\Delta_{\mathcal{V}_{1}}}{n},\quad h_M= \Big(n-\frac{1}{n}\Big)\frac{c_{M}}{24}+\frac{\Delta_{\mathcal{V}_{2}}}{n}\,.
\end{equation}
In what follows, we provide a short derivation in support of (\ref{Appendix}) for completeness. This will also constitute bulk of the pathway to compute symmetry resolved entropy in GCFT.

\subsection{Derivation of scaling dimension of composite twist operators for GCFT}
A very important ingredient in the calculation of the SREE in the Galilean theory would be the scaling dimension for composite twist operators in this theory. A similar computation for twist operators (without any charge) is provided in \cite{Basu:2015evh,Basu:2017aqn}. However, we need to revisit that to clarify some subtle issues and generalize it for our case in the presence of an Aharonov-Bohm flux. 
To go ahead, we need to start with the transformation rules of two stress-tensor $T_{1}$ and $T_{2}$ as mentioned in (\ref{T12}) under Galilean conformal transformations\footnote{General Galilean conformal transformations are given by $t \to  f(t)$ and $x \to x \partial_t f(t) + g(t)$, with free functions $f,g$.} \cite{Bagchi:2010vw},
\begin{align}
\begin{split}
\label{TE 1}
    &T_{1}(t^{\prime},x^{\prime})\rightarrow\Big(\frac{dt}{dt^{\prime}}\Big)^{2}T_{1}(t,x)+ 2\Big(\frac{dt}{dt^{\prime}}\Big)\Big(\frac{dx}{dt^{\prime}}\Big)T_{2}(t,x)+\frac{c_{L}}{12}\{t,t^{\prime}\}+ \frac{c_{M}}{12}\Big(\frac{dt}{dt^{\prime}}\Big)^{-1}\bigl[\bigl[(t,x),t^{\prime}\bigr]\bigr]\,,\\&
     T_{2}(t^{\prime},x^{\prime})\rightarrow\Big(\frac{dt}{dt^{\prime}}\Big)^{2}T_{2}(t,x)+\frac{c_{M}}{12}\{t,t^{\prime}\}\,.
\end{split}
\end{align}
Here $\{t,t^{\prime}\}$ denotes the Schwarzian derivative and is given by,
\begin{equation}
    \{t,t^{\prime}\}=\Bigg[\frac{d^{3}t}{dt^{\prime 3}}-\frac{3}{2}\Big(\frac{d^{2}t}{dt^{\prime 2}}\Big)^{2}\Big(\frac{dt}{dt^{\prime}}\Big)^{-1}\Bigg]\Big(\frac{dt}{dt^{\prime}}\Big)^{-1}\,.
\end{equation}
Also, $\bigl[\bigl[(t,x),t^{\prime}\bigr]\bigr]$ denotes the GCA Schwarzian, which is manifestly different from the CFT one, and given by the following formula,
\begin{equation}
    \bigl[\bigl[(t,x),t^{\prime}\bigr]\bigr]=\{(t,x),t^{\prime}\}-\frac{dx}{dt^{\prime}}\{t,t^{\prime}\}
\end{equation}
where\begin{equation}
    \{(t,x),t^{\prime}\}=\frac{d^{3}x}{dt^{\prime 3}} + 3\Bigg[\frac{1}{2}\Big(\frac{d^{2}t}{dt'^{2}}\Big)^{2}\frac{dx}{dt^{\prime}}\Big(\frac{dt}{dt^{\prime}}\Big)^{-1} -\Big(\frac{d^{2}t}{dt^{\prime 2}}\Big)\Big(\frac{d^{2}x}{dt^{\prime 2}}\Big)\Bigg]\Big(\frac{dt}{dt^{\prime}}\Big)^{-1}\,.
\end{equation} 

Now the conformal map between the n-sheeted Riemann surface $\mathcal{R}_{n,\alpha}$ and GCFT complex-plane can be achieved via the conformal transformations \cite{Bagchi:2014iea,Basu:2015evh},
\begin{equation}\label{CM 1}
    t=\Bigg(\frac{t^{\prime}-t_{1}}{t^{\prime}-t_{2}}\Bigg)^{\frac{1}{n}}
\end{equation}
and
\begin{equation}\label{CM 2}
    x=\frac{1}{n}\Bigg(\frac{t^{\prime}-t_{1}}{t^{\prime}-t_{2}}\Bigg)^{\frac{1}{n}}\Bigg(\frac{x^{\prime}-x_{1}}{t^{\prime}-t_{1}}-\frac{x^{\prime}-x_{2}}{t^{\prime}-t_{2}}\Bigg)\,.
\end{equation}

    Then \textcolor{black}{the} expectation value of the energy-momentum tensors  $T_{1}(t^{\prime},x^{\prime})$ and $T_{2}(t^{\prime},x^{\prime})$ in presence of flux $\alpha$ is given by the relation,
    \begin{align}
    \begin{split}\label{Flux1}
       & \langle T_{1}(t^{\prime},x^{\prime})\rangle_{\mathcal{R}_{n,\alpha}}=\Big(\frac{dt}{dt^{\prime}}\Big)^{2}\langle T_{1}(t,x)\rangle_{\mathcal{C}_{,\alpha}} + 2\Big(\frac{dt}{dt^{\prime}}\Big)\Big(\frac{dx}{dt^{\prime}}\Big)\langle T_{2}(t,x)\rangle_{\mathcal{C}_{,\alpha}}+\frac{c_{L}}{12}\{t,t^{\prime}\}
   + \frac{c_{M}}{12}\Big(\frac{dt}{dt^{\prime}}\Big)^{-1}\bigl[\bigl[(t,x),t^{\prime}\bigr]\bigr]\,,\\&
     \langle T_{2}(t^{\prime},x^{\prime})\rangle_{\mathcal{R}_{n,\alpha}}=\Big(\frac{dt}{dt^{\prime}}\Big)^{2}\langle T_{2}(t,x)\rangle_{\mathcal{C}_{,\alpha}}+\frac{c_{M}}{12}\{t,t^{\prime}\}\,.
    \end{split}
    \end{align}

       
    
Here we have used (\ref{TE 1}) to map $\mathcal{R}_{n,\alpha}$ to the complex plane $\mathcal{C_{,\alpha}}$ with a flux $\alpha$ inserted on it.   In the absence of flux $\alpha$, $\langle T_{1}(t,x)\rangle_{\mathcal{C}}$ and $\langle T_{2}(t,x)\rangle_{\mathcal{C}}$ vanishes for the ground state due to its symmetries, but in the presence of flux they are non-zero as argued in \cite{Goldstein:2017bua}. Then one-point functions $\langle T_{i}(t,x)\rangle_{\mathcal{C}_{,\alpha}}\,, i=1,2$ can be written as, $$\langle T_{i}\rangle_{\mathcal{C}_{,\alpha}}=\frac{\langle T_{i}\mathcal{V}_{1}\mathcal{V}_{2}\rangle_{\mathcal{C}}}
{\langle \mathcal{V}_{1}\mathcal{V}_{2}\rangle_{\mathcal{C}}}\,,$$ where scaling dimensions of the operators generating the flux are denoted by $\Delta_{\mathcal{V}_{1}}$ and $\Delta_{\mathcal{V}_{2}}$ respectively. They are primary fields and their \textcolor{black}{two-point functions} take the form mentioned in (\ref{GCFT2pt}). The three-point functions $\langle T_{i}\mathcal{T}_{\mathcal{\nu}}\mathcal{\bar{T}}_{\mathcal{\nu}}\rangle_{\mathcal{C}}$ can be computed using GCA ward identities \cite{Basu:2015evh} given below \footnote{Note that in \cite{Saha:2022gjw}, an intrinsic computation of $ \langle T_{1}\phi_{1}\phi_{2}\rangle$ and $ \langle T_{2}\phi_{1}\phi_{2}\rangle$ for CCFT are also provided. We can obtain the corresponding results for GCFT by interchanging $x$ and $t$ as mentioned in Appendix~(\ref{AppB}). Further, for us the sign conventions in these ward identities are slightly different from \cite{Basu:2015evh}, which we will consistently follow throughout.},
\begin{align}
\begin{split}
  &  \langle T_{1}\phi_{1}\phi_{2}\rangle=\sum_{i=1}^{2}\Bigg[\frac{1}{t-t_{i}}\partial_{t_{i}}-2h_{M}^{i}\frac{x-x_{i}}{(t-t_{i})^{3}} +\frac{1}{(t-t_{i})^{2}}(h_{L}^{i}-(x-x_{i})\partial_{x_{i}})\Bigg]\langle\phi_{1}\phi_{2}\rangle\,,\\&
   \langle T_{2}\phi_{1}\phi_{2}\rangle=\sum_{i=1}^{2}\Bigg[\frac{h_{M}^{i}}{(t-t_{i})^{2}}+\frac{1}{(t-t_{i})}\partial_{x_{i}}\Bigg]\langle\phi_{1}\phi_{2}\rangle\,,
\end{split}
\end{align}
where $\phi_1$ and $\phi_2$ are two primary operators with scaling dimensions $h_M^{(1,2)}$ and $h_L^{(1,2)}$ respectively and the two-point function $\langle \phi_1\phi_2\rangle$ is given in (\ref{GCFT2pt}). Then we get   
\begin{align}
     \begin{split}
  & \langle T_{1}(t,x)\rangle_{\mathcal{C}_{,\alpha}}=\Big(\frac{t(t_{1})-t(t_{2})}{(t(t^{\prime})-t(t_{1}))(t(t^{\prime})-t(t_{2}))}\Big)^{2}\\&\hspace{2.5cm}
    \Big [\Delta_{\mathcal{V}_{1}} + 2\Delta_{\mathcal{V}_{2}}\Big(\frac{x(x_{1},t_{1})-x(x_{2},t_{2})}{t(t_{1})-t(t_{2})}-\frac{x(x^{\prime},t^{\prime})-x(x_{1},t_{1})}{t(t^{\prime})-t(t_{1})}-\frac{x(x^{\prime},t^{\prime})-x(x_{2},t_{2})}{t(t^{\prime})-t(t_{2})}\Big)\Big]\,,\\&
      \langle T_{2}(t,x)\rangle_{\mathcal
      {C},\alpha}=
        \Delta_{\mathcal{V}_{2}}\Big(\frac{t(t_{1})-t(t_{2})}{(t(t^{\prime})-t(t_{1}))(t(t^{\prime})-t(t_{2}))}\Big)^{2}\,.
    \end{split}
   \end{align}
From (\ref{CM 2}), it follows that, $$t(t_{1})\rightarrow 0 ,t(t_{2})\rightarrow \infty, \quad \textrm{and}\quad  x(x_{1},t_{1})\rightarrow 0 ,x(x_{2},t_{2})\rightarrow\infty\,.$$
In this limit we get, 
\begin{align}
\begin{split}
     \langle T_{1}(t,x)\rangle_{\mathcal
      {C}_{,\alpha}}\sim \frac{1}{t(t^{\prime})^{2}}\Big[\Delta_{\mathcal{V}_{1}}-\frac{2\,\Delta_{\mathcal{V}_{2}}}{n}
\Big(\frac{x^{\prime}-x_{1}}{t^{\prime}-t_{1}}-\frac{x^{\prime}-x_{2}}{t^{\prime}-t_{2}}\Big)\Big]\,,\quad 
     \langle T_{2}(t,x)\rangle_{\mathcal
      {C}_{,\alpha}}\sim \frac{\Delta_{\mathcal{V}_{2}}}{t(t^{\prime})^{2}}\,.
\end{split}
\end{align}
Putting all the results together, we get from (\ref{Flux1}) 
\begin{align} 
 \begin{split} 
 \langle T_{1}(t^{\prime},x^{\prime})\rangle_{\mathcal{R}_{n,\alpha}}&=\Big(\frac{t_{12}}{t_{1}^{\prime}t_{2}^{\prime}}\Big)^{2}\Big[\Big\{\frac{\Delta_{\mathcal{V}_{1}}}{n^2}+\Big(1-\frac{1}{n^2}\Big)\frac{c_{L}}{24}\Big\}+  \\ &2\,\Big\{\frac{\Delta_{\mathcal{V}_{2}}}{n^2}+\Big(1-\frac{1}{n^2}\Big)\frac{c_{M}}{24}\Big\}\Big(\frac{x_{12}}{t_{12}}-\frac{x'-x_1}{t'-t_1}-\frac{x'-x_2}{t'-t_2}\Big)\Big]\,,\\
   \langle T_{2}(t^{\prime},x^{\prime})\rangle_{\mathcal{R}_{n,\alpha}}&=\Big(\frac{t_{12}}{t_{1}^{\prime}t_{2}^{\prime}}\Big)^{2}\Big[\frac{\Delta_{\mathcal{V}_{2}}}{n^2}+\Big(1-\frac{1}{n^2}\Big)\frac{c_{M}}{24}\Big]\,. \label{Final}
   \end{split}
\end{align}
A consistency check for the expression given in (\ref{Final}) has been provided in Appendix~(\ref{AppA}) by taking a non-relativistic limit of two copies of the CFT stress tensor. Now from (\ref{Final}) we can read off the scaling dimensions of the composite twist operators as mentioned in
(\ref{Appendix}). 

\subsection{SRRE for GCFT:}
Then we can move on to compute $Z_n(\alpha)$, the partition function on $\mathcal{R}_{n,\alpha}$ as defined in (\ref{fluxpartition}). As mentioned earlier this comes from the two point correlator of the twist field. This is given in (\ref{GCFT2pt}). Setting the UV cutoff to unity, we get:  
\begin{equation} \label{compare}
\textcolor{black}{Z}_{n}(\alpha)=t_{12}^{-2\left[\frac{\Delta_{\mathcal{V}_{1}}}{n}+(n-\frac{1}{n})\frac{c_{L}}{24}\right]} \exp\Big[-2\Big\{\frac{\Delta_{\mathcal{V}_{2}}}{n}+
\Big(n-\frac{1}{n}\Big)\frac{c_{M}}{24}\Big\}\frac{x_{12}}{t_{12}}\Big]\,.
\end{equation}
This precisely matches with (\ref{compare1}) which was obtained from the CFT result in the limit of zero boost $\beta\rightarrow 0$. Thus, it is obvious that the expression for SREE in GCFT coming from this intrinsic computation will match with what we get by taking the non-relativistic limit of the CFT result.
\medskip

This result can be extended easily for a system at finite temperature and/or finite spatial extent. We will discuss these finite temperature results mainly from the point of view of  Carrollian theory (CCFT) in the Sec.~(\ref{sec6}) due to the obvious holographic implications. The corresponding results for GCFT can be obtained by interchanging the spatial and temporal interval. As we will see, they reduce to the above expression in the zero temperature or infinite system size limit. 



\section{CFT to BMS: Symmetry Resolution }\label{sec6}

Finally, we would like to explore the ultra-relativistic limit of the symmetry-resolved entropy. The ultra-relativistic or Carrollian limit is crucial in light of holographic duality for gravity in asymptotically flat spacetime \cite{flatHol,flatHol1,flatHol2,flatHol3,flatHol4,flatHol7}, as already hinted at in the Introduction. One of the easiest ways to think about Carrollian spacetime is by starting with Minkowski and then taking the speed of light to zero \cite{Levy1965,sengu,henne}. Even before putting any dynamical consideration (particles of fields on such a space-time), one notices that the metric becomes degenerate, and as a result, the notion of causality becomes drastically different. The lightcone in the Minkowski space-time shrinks to a line along the temporal axis in this limit. This is behind the consensus that particles (massless or massive) in a Carroll space-time do not move, giving rise to the concept of \textit{ultra-locality}. In the simplest field theory context of a massless scalar field, the Hamiltonian in Minkowski spacetime is:
\begin{eqnarray} \label{hamphi}
    H \sim \int \pi^2 + c^2 (\nabla \phi)^2. 
\end{eqnarray}
Due to the positive definiteness of the gradient term in the Hamiltonian, minimizing the energy would mean a strong correlation between fields at nearby points. That essentially means that a zero-temperature theory has high subsystem entanglement. This is one of the ways to understand why unitarity and locality force ground states in relativistic field theories to be strongly entangled. Taking $c \rightarrow 0$ in \eqref{hamphi}, on the other hand, relaxes the above condition almost to triviality, and one can readily argue about very low-entangled ground states in a Carrollian theory \cite{Banerjee:2023jpi} where the momentum term dominates. 
\medskip

 In fact, from a generic symmetry argument, EE (for a subsystem A) was computed in Carrollian conformal field theory (CCFT) to be of the form: \cite{Bagchi:2014iea, Grumiller:2019xna,Jiang:2017ecm}:
\begin{eqnarray} \label{EE_CCFT}
		S (A) = \frac{c_L}{6} \ln \left( \frac{x_{12}}{\varepsilon_1}\right) + \frac{c_{\textcolor{black}{M}}}{6} \left( \frac{t_{12}}{x_{12}} - \frac{\varepsilon_2}{\varepsilon_1}\right)
	\end{eqnarray}
where $x_{12}$ and $t_{12}$ respectively are the spatial separation and temporal separation of two points in a two-dimensional flat Carroll manifold, and the $\varepsilon_i$ are the regulators. When proposed as a dual to gravity in 3D flat space-time, $c_L = 0, c_M = 3/(2G)$, where the gravity theory is pure Einstein gravity. Note that, when derived as a limit of relativistic CFTs with zero chiral anomaly, $c = \bar{c}$, vanishing of $c_L$ is automatically guaranteed (see Appendix \ref{AppB}). Evidently, from \eqref{EE_CCFT}, we see that for an equal time interval, CCFTs dual to 3D Einstein gravity don't have any subsystem entanglement at all. This validates the argument, made intuitively and presented in the preceding paragraph. 
\medskip

This can be further understood from the ultra-local feature of Carrollian field theories in general (not necessarily conformal). Ultra-locality forces energy eigenstates to be localized in space, as expected from the curious causal structure of Carroll physics. Hence, the ground state and any stationary state are a direct product of local Hilbert space states having vanishing entanglement \cite{Ara:2023pnn, Banerjee:2023jpi}. There, however, are a couple of ways to see entanglement in a CCFT. One is by breaking the chirality and hence $c_L \neq 0$. On the gravity side, this is achieved by a topologically massive term \cite{Bagchi:2012yk}. On the field theory side, the same can be done by a non-trivial Aharonov-Bohm flux \cite{tovmasyan2018preformed}. The second route, which we take here, is to go to a boosted interval $t_{12} \neq 0$, even if $c_L = 0$. This is one of the reasons why we should keep the SREE results in a boosted frame for ordinary CFTs handy in Sec.~(\ref{sec3}).

\subsection{Computation at zero temperature}
We start with (\ref{sboost}) and again continue to the Lorenztian time. We write our coordinates as $z_{12}=L e^{\psi}=L\cosh(\psi)+L\sinh(\psi),$ where $\psi$ is related to the boost parameter. 
Then, we can define, 
\begin{equation} \label{param} \beta=\coth{\psi},\quad \gamma=\sinh{\psi},\quad \gamma \beta= \cosh{\psi}\,.\end{equation} 
Note that this is a somewhat different identification than we made before, since this limit is more subtle. We can identify, $\gamma\beta x_{12}=L\cosh{\psi},\gamma t_{12}= L\sinh{\psi}$ and this gives, 
\begin{equation} \label{param1} z_{12}=\gamma(\beta x_{12}+t_{12}),\quad \bar{z}_{12}=\gamma(\beta x_{12}-t_{12})\,.\end{equation} 
Now (\ref{sboost}) can be written using these identifications,
\begin{align}
    \begin{split} \label{scarroll}
&\textcolor{black}{Z}_{n}(\alpha)=\Big[\frac{\gamma\beta x_{12}}{\varepsilon}\left(1+\frac{t_{12}}{\beta x_{12}}\right)\Big]^{\Big\{-\frac{c}{12}(n-\frac{1}{n})-\frac{2}{n}\Delta_\mathcal{V}\}\Big\}}\Big[\frac{\gamma\beta x_{12}}{\varepsilon}\left(1-\frac{t_{12}}{\beta x_{12}}\right)\Big]^\Big{\{-\frac{\bar{c}}{12}(n-\frac{1}{n})-\frac{2}{n}\bar{\Delta}_\mathcal{V}\Big\}}
\end{split}
    \end{align}
Recall that the Carrollian limit entails taking the speed of light to zero, hence in this limit, the Lorentz boost parameter goes to infinity \footnote{For explicit construction of Carroll Conformal scalar field theories from relativistic theories via infinite boost mechanism, one can look at \cite{Bagchi:2022nvj}. A description of unequal central charges appearing in this setup can be found in \cite{2024arXiv240116482B}.}; 
\begin{equation} \label{param2} \beta\rightarrow\infty,\quad \frac{1}{\beta}=\tanh{\psi}\rightarrow 0,\quad \gamma\beta=\cosh{\psi}\rightarrow 1\,. \end{equation} 
Setting the UV cutoff $\varepsilon$ to unity, we get by reorganizing (\ref{scarroll}):
\begin{align}
    \begin{split} \label{scarroll1}
    \textcolor{black}{Z}_{n}(\alpha)=(x_{12})^{-\Big[\frac{c+\bar{c}}{12}(n-\frac{1}{n})+\frac{2}{n}(\Delta_\mathcal{V}+\bar{\Delta}_{\mathcal{V}})\Big]}
\Big(1+\frac{1}{\beta}\frac{t_{12}}{x_{12}}\Big)^{-\Big[\frac{c}{12}(n-\frac{1}{n})+\frac{2}{n}\Delta_\mathcal{V}\Big]}\Big(1-\frac{1}{\beta}\frac{t_{12}}{x_{12}}\Big)^{-\Big[\frac{\bar{c}}{12}(n-\frac{1}{n})+\frac{2}{n}\bar{\Delta}_\mathcal{V}\Big]}\,.\end{split}
    \end{align}
In $\frac{1}{\beta}\rightarrow 0$  limit, assuming $c,\bar c$ to be very large as in the Galilean case, we can approximate (\ref{scarroll1}) as 
\begin{align}
    \begin{split} \label{scarroll2}
\textcolor{black}{Z}_{n}(\alpha)=(x_{12})^{-\Big[\frac{c+\bar{c}}{12}(n-\frac{1}{n})+\frac{2}{n}(\Delta_\mathcal{V}+\bar{\Delta}_{\mathcal{V}})\Big]}\exp\Big[-\Big\{\frac{c-\bar{c}}{12\,\beta}\Big(n-\frac{1}{n}\Big)+\frac{2}{n\,\beta}(\Delta_{\mathcal{V}}-\bar{\Delta}_{\mathcal{V}})\Big\}\frac{t_{12}}{x_{12}}\Big]\,.
\end{split}
    \end{align}

Now, note that we started taking this limit from the CFT result in (\ref{sboost}) for the boosted interval. Furthermore, comparing the expressions for $z_{12}$ and $\bar{z}_{12}$ for the GCFT and CCFT boosted intervals, we can easily see that they are related via $\bar{z}_{12} \to -\bar{z}_{12}$.
Therefore, to make sense of the ultra-relativistic limit, we have to replace $\bar{c}$ by $-\bar{c}$ and $\bar{\Delta}_{\mathcal{V}}$ by $-\bar{\Delta}_{\mathcal{V}}$, which actually follows from the choice of the representation of the BMS$_3$ (or Carrollian) generators as discussed in Appendix~(\ref{AppB}). Then we can identify the following combination of central charges: 
\begin{equation}
\label{BMSident} c-\bar{c}=c_{L},\quad \frac{1}{\beta}(c+\bar{c})=c_{M}\end{equation} and also denoting \footnote{For an example of vertex operators coming from BMS invariant scalar theory readers are referred to \cite{Hao:2021urq}.} 
$$\Delta_{\mathcal{V}_{1}}=\Delta_{\mathcal{V}}-\bar{\Delta}_{\mathcal{V}}\quad \Delta_{\mathcal{V}_{2}}=\frac{1}{\beta}(\Delta_{\mathcal{V}}+\bar{\Delta}_{\mathcal{V}})\,,$$ 
we finally get, 
\begin{align}
    \begin{split} \label{scarroll3}
\textcolor{black}{Z}_{n}(\alpha)= {x_{12}}^{-[\frac{c_L}{12}(n-\frac{1}{n})+\frac{2\,\Delta_{\mathcal{V}_{1}}}{n}]} \exp\Bigg[-\Big\{\frac{c_M}{12}\Big(n-\frac{1}{n}\Big)+\frac{2\,\Delta_{\mathcal{V}_{2}}}{n}\Big\}\frac{t_{12}}{x_{12}}\Bigg]\,.
\end{split}
    \end{align}
    Note that the structure of (\ref{scarroll3}) is the same as that of (\ref{sgcft}). In fact, (\ref{scarroll3}) can be obtained from (\ref{sgcft}) by swiftly interchanging $x_{12}$ and $t_{12}$. This has also been observed when computing EE in \cite{Bagchi:2014iea,Basu:2015evh}, which isn't surprising in the sense such a duality in two dimensions under the exchange of base and fibre directions are well expected. Again we note that, commensurate with this $x \leftrightarrow t$ interchange, $\frac{1}{\beta}$ itself becomes the contraction parameter for our case. These should again be compared with the Appendix \eqref{AppB}.
    \medskip
    
    Then following the same procedure discussed in Sec.~(\ref{sec4}), and using the vertex operators mentioned in (\ref{vert}), we obtain the SREE of the ground state for BMS$_3$ as follows:
  \begin{align}
\begin{split} \label{SREEBMS}
S_{\textrm{BMS}}(N_{A})=\frac{c_{L}}{6}\ln{x_{12}}+\frac{c_{M}}{6}\frac{t_{12}}{x_{12}}-\frac{1}{2}+\frac{1}{2} \ln \left(\frac{\pi }{L'}\right)\,,
\end{split}
\end{align}
where we have again assumed the interval $K_{1}\ln x_{12}+K_{2}\frac{t_{12}}{x_{12}}= L'$ is $\gg 1$ and we have kept only the leading order terms. The form of (\ref{SREEBMS}), as we saw before, is same as that of (\ref{SREEGCFT}) if one interchanges $x$ and $t\,.$  An interesting point to note, for the theory dual to Einstein gravity in 3d flat space ($c_L=0$), even if we consider an equal-time interval ($t_{12}=0)$ at the boundary  $S_{\textrm{BMS}}(N_{A})$ is non zero ($K_1\neq 0$). Although the leading term in (\ref{SREEBMS}) vanishes, there will be a $\mathcal{O}(1)$ term and a subleading (in $L'$) term remaining\footnote{One should also note for Einstein gravity with unboosted interval, (\ref{scarroll3}) just becomes one, hence the probability becomes a subsystem localized delta function.}. It takes the same form as shown in (\ref{SREEGCFT}), and we have 
\begin{align}
\begin{split} \label{SREEBMS1}
S_{\textrm{BMS}}(N_{A})\Big|_{c_L=0, t_{12}=0}=\frac{1}{2} \ln \left(\frac{\pi }{L'}\right)-\frac{1}{2}\,.
\end{split}
\end{align}
Now, as in (\ref{TotalEntropyGCFT}), we will see that the total entanglement will vanish which corresponds to the known result \cite{Basu:2015evh,Basu:2017aqn} for CCFT. Although the total von Neumann entropy for an equal-time interval can vanish, the entropy associated with each of the charge sectors can be non-vanishing (in the subleading terms).
Note that, the above expression will contain information about the scaling dimensions of the vertex operators generating the flux. 

\subsection{Finite temperature/Finite spatial extent:}
We will now extend our results for $S_{n}(\alpha)$ mentioned in (\ref{scarroll2}) for a system at finite temperature  and/or finite spatial extent. To avoid any confusion with the Lorentz boost parameter $\beta$, let us set our inverse temperature notation as $T=\beta_{T}^{-1}$, which we will follow throughout the section. The result obtained for the CCFT computation can be extended to GCFT by suitably interchanging the role of the spatial and the temporal intervals, as mentioned earlier.  We start with the two-point function of the composite twist operators after mapping the $n$-sheeted Riemann surface to the plane for CCFT. 
\begin{equation} \label{GCFT2pt1a}
\langle\mathcal{T}_{\nu}\mathcal{\tilde{T}}_{\nu}\rangle=\Big(\frac{x_{12}}{\varepsilon}\Big)^{-2h_{L}} \exp\Big(-2h_{M}\frac{t_{12}}{x_{12}}\Big)\,, 
\end{equation}
with $$h_{L}=\Big(n-\frac{1}{n}\Big)\frac{c_{L}}{24}+\frac{\Delta_{\mathcal{V}_{1}}}{n}\, \quad  h_{M}=\Big(n-\frac{1}{n}\Big)\frac{c_{M}}{24}+\frac{\Delta_{\mathcal{V}_{2}}}{n}\, \quad \Delta_{\mathcal{V}_{1}}=\frac{1}{2}\Big(\frac{\alpha}{2\pi}\Big)^{2} K_1,\quad \Delta_{\mathcal{V}_{2}}=\frac{1}{2}\Big(\frac{\alpha}{2\pi}\Big)^{2}K_2\,.$$
Now to achieve the result for finite temperature theory we have to further map this correlator from the plane to a thermal cylinder \cite{Calabrese:2004eu,Calabrese:2009qy}. For our case, this can be achieved by the following transformation \cite{Basu:2015evh} \footnote{The zero temperature version of this (null) plane to (null) cylinder map is given by, $x = e^{i\xi}, t = i \tau e^{i\xi}$.}, 
\begin{equation}
x=e^{2\pi\xi/\beta_{T}},\quad t=\frac{2\pi\tau}{\beta_{T}}e^{2\pi\xi/\beta_{T}}\,.
\end{equation}
Here, $\xi$ and $\tau$ are the coordinates on the cylinder. One can easily see that through this coordinate change we have essentially compactified one of the directions. Then we obtain (after suitably rescaling the UV-cutoff \cite{Calabrese:2004eu,Calabrese:2009qy}), the two point function becomes:
\begin{align} 
\begin{split}\label{GCFT2pt1b}
\langle\mathcal{T}_{\nu}(\xi_1,\tau_1)\mathcal{\tilde{T}}_{\nu}(\xi_2,\tau_2)\rangle&=e^{\frac{2\pi}{\beta_{T}}(h_L(\xi_1+\xi_2)+h_M(\tau_1+\tau_2)}\Big(\frac{x_{12}}{\varepsilon}\Big)^{-2h_{L}} \langle\mathcal{T}_{\nu}(x_1(\xi_1),t_1(\xi_1,\tau_1))\mathcal{\tilde{T}}_{\nu}(x_2(\xi_2),t_2(\xi_2,\tau_2))\rangle\,, \\&
=\Big[\frac{\beta_{T}}{\pi\varepsilon}\sinh\Big(\frac{\pi\xi_{12}}{\beta_{T}}\Big)\Big]^{-2\,h_{L}}\exp\Big[-\frac{2\,h_M\pi\tau_{12}}{\beta_{T}}\coth\Big(\frac{\pi\xi_{12}}{\beta_{T}}\Big)\Big]\,.
\end{split}
\end{align}
Consequently, we can find, 
\begin{align} 
\begin{split}
\textcolor{black}{Z}_{n}(\alpha)\sim \langle\mathcal{T}_{\nu}(\xi_1,\tau_1)\mathcal{\tilde{T}}_{\nu}(\xi_2,\tau_2)\rangle &= \Big[\frac{\beta_{T}}{\pi\varepsilon}\sinh\Big(\frac{\pi\xi_{12}}{\beta_{T}}\Big)\Big]^{-2\,h_{L}}\exp\Big[-\frac{2\,h_M\pi\tau_{12}}{\beta}\coth\Big(\frac{\pi\xi_{12}}{\beta_{T}}\Big)\Big]\\&
=\textcolor{black}{Z}_{n}(\alpha=0)e^{-\frac{\alpha^2\,\bar{L}}{4\pi^2\,n}}\,,
\end{split}
\end{align}
where we have used, 
\begin{align} 
\begin{split}
&\textcolor{black}{Z}_{n}(\alpha=0)=\exp\Bigg[-\frac{\pi\,c_M\,(n-\frac{1}{n}) \tau_{12} \coth \left(\frac{\pi\, \xi_{12}}{\beta_{T} }\right)}{12\, \beta_{T} }\Bigg]\Bigg(\frac{\beta_{T}  \sinh \left(\frac{\pi\, \xi_{12}}{\beta_{T} }\right)}{\varepsilon\,\pi}\Bigg)^{-\frac{1}{12} c_L \left(n-\frac{1}{n}\right)}\,,\\& \bar{L}=K_{1} \ln \Bigg(\frac{\beta_{T}  \sinh \left(\frac{\pi \xi_{12}}{\beta_{T} }\right)}{\pi\,\varepsilon}\Bigg)+\frac{\pi\,K_{2} \left(\tau_{12} \coth \left(\frac{\pi\,\xi_{12}}{\beta_{T} }\right)\right)}{\beta_{T} }\,.\end{split}
\end{align}
Then using (\ref{fluxpartition}) we get, 
\begin{align} 
\begin{split}
\textcolor{black}{\mathcal{Z}_n}(N_{A})=\frac{\textcolor{black}{Z}_{n}(\alpha=0)}{2\pi}\sqrt{\frac{4\pi^{3}\,n}{\bar{L}}} e^{-\frac{\pi^{2}N_{A}^{2}n}{\bar{L}}}\,.
\end{split}
\end{align}
Finally we get the SREE for CCFT at finite temperature as given below, 
\begin{align}
\begin{split} \label{SREEGCFTFinitetemp}
S_{\textrm{BMS}}(N_{A})\Big|_{\beta_{T}}=\frac{c_L}{6}\, \ln \Bigg(\frac{\beta_{T}  \sinh \left(\frac{\pi \, \xi_{12}}{\beta_{T} }\right)}{\pi  \,\varepsilon}\Bigg)+\frac{c_M}{6}\,\frac{\pi\,\tau_{12} \coth \left(\frac{\pi \,\xi_{12}}{\beta_{T}}\right)}{\beta_{T} }-\frac{1}{2}+\frac{1}{2} \ln \left(\frac{\pi }{\bar{L}}\right)\,.
\end{split}
\end{align}
Again we can easily see that, using (\ref{fluxpartition1}) we can correctly produce the total EE at finite temperature:
\begin{align}
\begin{split} \label{FiniteTemp}
S_{\textrm{BMS}}\Big|_{\beta_{T}}=\frac{c_L}{6}\, \log \Bigg(\frac{\beta_{T}  \sinh \left(\frac{\pi \, \xi_{12}}{\beta_{T} }\right)}{\pi  \,\varepsilon}\Bigg)+\frac{c_M}{6}\,\frac{\pi\,\tau_{12} \coth \left(\frac{\pi \,\xi_{12}}{\beta_{T} }\right)}{\beta_{T} }\,.
\end{split}
\end{align}
This straightforwardly matches the results obtained in \cite{Basu:2015evh}. Moreover, we can check that, for low temperature limit i.e. $\beta_{T} \rightarrow \infty,$ we get back our results from previous sections:
\begin{align}
\begin{split}
&\bar{L}\approx L'= K_{1}\ln x_{12}+K_{2}\frac{t_{12}}{x_{12}}\,, \\& \frac{c_L}{6}\, \ln \Bigg(\frac{\beta_{T}  \sinh \left(\frac{\pi \, \xi_{12}}{\beta_{T} }\right)}{\pi  \,\varepsilon}\Bigg)+\frac{c_M}{6}\,\frac{\pi\,\tau_{12} \coth \left(\frac{\pi \,\xi_{12}}{\beta_{T} }\right)}{\beta_{T} }\approx \frac{c_{L}}{6}\ln{x_{12}}+\frac{c_{M}}{6}\frac{t_{12}}{x_{12}}\,.
\end{split}
\end{align}
Using these, we can easily verify, after identifying $\tau_{12}\sim t_{12}$ and $\xi_{12}\sim x_{12},$ that (\ref{SREEGCFTFinitetemp}) reduces to the zero-temperature case as shown in (\ref{SREEBMS}) correctly.
\medskip

A similar result can be obtained for a system with finite spatial size $L_{\beta}\,$ (not to be confused with $\bar{L}$). In that case, we simply have to replace $\beta_{T}$ by $i\,L_{\beta}$ in (\ref{FiniteTemp}) following the idea of \cite{Calabrese:2004eu} \footnote{Finally, we can also obtain the corresponding result for GCFT by interchanging the role of $\xi$ and $\tau$ as discussed previously.}. The the result for finite spatial extent looks like:
\begin{align}
\begin{split} \label{SREEGCFTFinitespatial}
S_{\textrm{BMS}}(N_{A})\Big|_{L_{\beta}}=\frac{c_L}{6}\, \log \Bigg(\frac{L_{\beta} \sin \left(\frac{\pi \, \xi_{12}}{L_{\beta}}\right)}{\pi  \,\varepsilon}\Bigg)+\frac{c_M}{6}\,\frac{\pi\,\tau_{12} \cot \left(\frac{\pi \,\xi_{12}}{L_{\beta}}\right)}{L_{\beta}}-\frac{1}{2}+\frac{1}{2} \log \left(\frac{\pi }{\bar{L}}\right)\,,
\end{split}
\end{align}
with, $$\bar{L}=K_{1} \log \Bigg(\frac{L_{\beta} \sin \left(\frac{\pi \xi_{12}}{L_{\beta}}\right)}{\pi\,\varepsilon}\Bigg)+\frac{\pi\,K_{2} \left(\tau_{12} \cot \left(\frac{\pi\,\xi_{12}}{L_{\beta}}\right)\right)}{L_{\beta}}\,.$$
Again to get the total EE we make use of (\ref{fluxpartition1}) and compute the sum of $S^c$ and $S^f$ as we did earlier, which readily gives the total EE:
\begin{align}
\begin{split}\label{TotalEntropyCCFTa}
S_{\textrm{BMS}}(A)\Big|_{L_{\beta}}=&S^c+S^f,\\&
=\frac{c_L}{6}\, \log \Bigg(\frac{L_{\beta} \sin \left(\frac{\pi \, \xi_{12}}{L_{\beta}}\right)}{\pi  \,\varepsilon}\Bigg)+\frac{c_M}{6}\,\frac{\pi\,\tau_{12} \cot \left(\frac{\pi \,\xi_{12}}{L_{\beta}}\right)}{L_{\beta}}\,.
\end{split}
\end{align}

The final result matches the previously obtained results of \cite{Bagchi:2014iea, Basu:2015evh}.
\medskip

Again we can see from (\ref{SREEGCFTFinitetemp}) and (\ref{SREEGCFTFinitespatial}) that, even for Einstein gravity ($c_L=0$) and an equal $\tau$ interval we can have non-vanishing SREE terms (assuming $K_1 \neq 0$).  But still we get vanishing total EE in this setting, as expected from earlier literature. Although we will defer a full-fledged analysis of SREE intrinsically performed from the flat space version of holography \cite{Bagchi:2010zz} to a future work, but it is tempting to conjecture that the result (\ref{SREEGCFTFinitespatial}) can be thought of as the result for dual field theory living at the boundary of the Global Flat Space (GFS). It can be intuitively seen that this is true, as one has to typically map the result of the EE from the plane to a cylinder to match what comes from the holographic analysis in the background of GFS.

\section{Symmetry Resolution for BMS$_{3}$-Kac-Moody }\label{sec7}
Although  BMS$_{3}$ is equivalent to the group of conformal isometries of two-dimensional flat Carroll space-time, the most natural way of its appearance is through asymptotic symmetries of asymptotically flat space-time \cite{Bagchi:2010zz}. At the null infinity of asymptotically flat spacetime, the finite dimensional Poincaré isometry group is enhanced to the infinite dimensional BMS$_3$ group. Hence, it's crucial to know the holographic origin of SREE for Carrollian theories discussed in the last section. In this section, we attempt to make this idea concrete.
\medskip 

Over and above the Ryu-Takayanagi's holographic correspondence for EE \cite{Ryu:2006bv,Ryu:2006ef}, the additional structure required for symmetry resolution is a dual source term for the $U(1)$ charge operator $e^{i \alpha \hat{N}_A}$ appearing in \eqref{fluxpartition}. A straightforward way to incorporate this is a $U(1)$ Wilson line operator for each chiral half of the CFT, inserted in the bulk, whose end points are anchored at the boundary points of the boundary subsystem. This calls for coupling a $U(1) \times U(1)$ Chern-Simons theory with levels $k, \Bar{k}$, with the bulk gravity. The asymptotic symmetry algebra for this theory on asymptotically AdS$_3$ background is  $\mathfrak{vir} \oplus \mathfrak{u}(1)_k \oplus \mathfrak{u}(1)_{\bar{k}}$. 

\subsection{Results from Virasoro-Kac-Moody}
Let us start with the Lorentzian calculation first. We would not be having a full-fledged intrinsically flat space version of this holographic computation of SREE, but take a limiting viewpoint later in the section.  Our starting point, as discussed previously, is the analysis of AdS$_3$ with two copies of $U(1)$ Chern-Simons theory as in \cite{Zhao:2020qmn} and then carefully taking the limit of the AdS radius $l \rightarrow \infty$. For completeness, we will first revisit and present the results discussed in \cite{Zhao:2020qmn}, before taking the Carrollian limit relevant to a holographic connection.
\medskip

Before proceeding further, we would now like to remind the readers that following (\ref{fluxpartition2}-\ref{fluxpartition1}) the SRRE takes the following form when we use the so called generating function method, as outlined in \cite{Capizzi_2020,Zhao:2020qmn}: 
\begin{align}
\begin{split}
S_{n}(N_A)= S_n(A) +\frac{1}{1-n}\ln\frac{P_n(N_A)}{P_1(N_A)^n}\,,
\end{split}
\end{align}
where $S_n$ is the total RE and $P_1(N_A)$ denotes the probability of having $N_A$ number of particles in the subregion $A$. Also, by definition, we have:

\begin{equation} \label{chargemoment}
    P_n(N_A)=\int_{-\pi}^{\pi}\frac{d\alpha}{2\pi}e^{-i\,\alpha\,N_A}\,\Bigg[\frac{\textrm{Tr}({\hat{\rho}}^n_A e^{i\,\alpha\,{\hat{N}}_A})}{\textrm{Tr}({\hat{\rho}}^n_A)}\Bigg]\,.\end{equation}  The SREE ($S(N_A)$) is then given by, 
\begin{align}
\begin{split} \label{Stot}
S(N_A)= S(A) +\lim_{n\rightarrow 1}\frac{1}{1-n}\ln\frac{P_n(N_A)}{P_1(N_A)^n}\,.
\end{split}
\end{align}
Note that in these set of definitions all the charge dependence is shifted only to the second term. 
Now for Virasoro-Kac-Moody, $S_n(A)$ as well as $S(A)$ for a boosted interval is again given by (\ref{Renyi}) and (\ref{boostEE}) respectively. In what follows, to compute the $P_n(N_A)$ we will follow the approach taken by \cite{Zhao:2020qmn}. 
\medskip

It can be shown that the operator $e^{i\,\alpha\,\hat{N}_A}$ inserted into the Riemann surface for the computation of the charged partition function, as shown in (\ref{fluxpartition}), is nothing but the product of the vertex operators $\mathcal{V}\mathcal{V}^{\dagger}$ \cite{Goldstein:2017bua}. Now, the full vertex operator $\mathcal{V}(z,\bar{z})$ is product of holomorphic and anti-holomorphic part i.e. $\mathcal{V}(z,\bar{z})=\mathcal{V}(z)\mathcal{\bar{V}}(\bar{z})$ \footnote{In \cite{Zhao:2020qmn} they take the following form: $\mathcal{V}(z)=:\exp\Big(\frac{\alpha}{2\pi}\int^z dw \hat{J}(w)\Big):$\,, where, $\hat{J}(w)$ is the Kac-Moody current. Similarly for $\mathcal{\bar{V}}(\bar{z})\,.$}. They are both primary operators with respect to the Virasoro algebra and the $U(1)$ part. Their conformal dimensions are given by,
\begin{align}
\begin{split}
   \Delta_{\mathcal{V}}=\frac{1}{2}\Big(\frac{\alpha}{2\pi}\Big)^{2} \frac{k}{2},\quad \bar{\Delta}_{\mathcal{V}} =\frac{1}{2}\Big(\frac{\alpha}{2\pi}\Big)^{2} \frac{\bar{k}}{2}\,.
\end{split}
\end{align}
Where $k,\bar{k}$ are the levels of the two $U(1)$s.
Next we compute $P_n(N_A)$, noting that this can be done by computing the total charge at the subregion in the following way, 
\begin{align}
\begin{split} \label{chargemoment1}
\langle i\,\textcolor{black}{{\hat{N}}_A}\rangle_{n}=\textcolor{black}{\frac{\textrm{Tr}(i\,{\hat{N}}_A\,{\hat{\rho}}^n_A e^{i\,\alpha\,{\hat{N}}_A})}{\textrm{Tr}({\hat{\rho}}^n_A\,e^{i\,\alpha\,{\hat{N}}_A})}}=\frac{\partial}{\partial \alpha}\ln\Big(\frac{\textrm{Tr}({\hat{\rho}}^n_A e^{i\,\alpha\,{\hat{N}}_A})}{\textrm{Tr}({\hat{\rho}}^n_A)}\Big)\,.
\end{split}
\end{align}
Now for the Virasoro-Kac-Moody theory, we can explicitly find, 
\begin{align}
\begin{split}
\langle i\,\textcolor{black}{{\hat{N}}_A}\rangle_{n}= -\frac{k\,\alpha}{4\pi^2\,n}\ln\Big(\frac{z_{12}}{\varepsilon}\Big)-\frac{\bar{k}\,\alpha}{4\pi^2\,n}\ln\Big(\frac{\bar{z}_{12}}{\varepsilon}\Big)\,.
\end{split}
\end{align}
Then from (\ref{chargemoment1}) it follows, 
\begin{align}
\begin{split}
\Big(\frac{\textrm{Tr}(\textcolor{black}{{\hat{\rho}}}^n_A e^{i\,\alpha\,\hat{N_A}})}{\textrm{Tr}(\textcolor{black}{{\hat{\rho}}}^n_A)}\Big)&=\exp\Bigg[-\frac{1}{4\pi^2\,n}\int_0^{\alpha}\alpha' \Big\{k\,\ln\Big(\frac{z_{12}}{\varepsilon}\Big)+\bar{k}\,\ln\Big(\frac{\bar{z}_{12}}{\varepsilon}\Big)\Big\}d\alpha'\Bigg]\,,\\&=\Big(\frac{z_{12}}{\varepsilon}\Big)^{-\frac{k}{2n}(\frac{\alpha}{2\pi})^2}\Big(\frac{\bar{z}_{12}}{\varepsilon}\Big)^{-\frac{\bar{k}}{2n}(\frac{\alpha}{2\pi})^2}\,.
\end{split}
\end{align}
At this point, we set the UV-cutoff ($\varepsilon$) to unity. Then from (\ref{chargemoment}) we use a saddle-point approximation in the spirit of earlier sections (i.e. using the large interval limit, $z_{12}, \bar{z}_{12} \gg 1$) to extend the integration range and then preforming a Gaussian integration, resulting in:
\begin{align}\label{prob}
\begin{split}
P_n(N_A)&=\int_{-\infty}^{\infty}\frac{d\alpha}{2\pi}\exp\Big[-i\,\alpha\,N_A-\frac{k}{2n}\Big(\frac{\alpha}{2\pi}\Big)^2\ln z_{12}-\frac{\bar{k}}{2n}\Big(\frac{\alpha}{2\pi}\Big)^2\ln \bar{z}_{12}\Big],\\&
=\sqrt{\frac{2\pi\,n}{k\ln z_{12}+\bar{k}\ln \bar{z}_{12}}}e^{-\frac{2\pi^2\,n\,N^2_A}{k\ln z_{12}+\bar{k}\ln \bar{z}_{12}}}\,.
\end{split}
\end{align}
Finally, by using (\ref{Stot}) we get the final from of the SREE:
\begin{align}
\begin{split} \label{CFTKacMoody}
S(N_A)=\frac{c}{6}\ln z_{12}+\frac{\bar{c}}{6}\ln\bar{z}_{12}-\frac{1}{2}\ln\Big(\frac{k \ln z_{12}+\bar{k}\ln \bar{z}_{12}}{2\pi}\Big)-\frac{1}{2}\,.
\end{split}
\end{align}

\subsection*{{\bf A consistency check:}}
We carry out the consistency check on the result \eqref{CFTKacMoody} originally presented in \cite{Zhao:2020qmn}. From (\ref{prob}) we can calculate the probability of finding $N_{A}$ number of particles in the subregion A 
\begin{equation} \label{Gaussian}
    P_{1}(N_{A})=\sqrt{\frac{2\pi }{k\ln{z_{12}+\Bar{k}\ln{\Bar{z}_{12}}}}} e^{-\frac{-2\pi^{2}N_{A}^{2}}{k\ln{z_{12}+\Bar{k}\ln{\Bar{z}_{12}}}}}\,.
\end{equation}\\
Now following (\ref{fluxpartition1}), the configurational entropy 
comes out to be,
\begin{equation} \label{conf}
    S^{c}=\int_{-\infty}^{\infty}S(N_{A}) P_{1}(N_{A})\,dN_{A}=\frac{c}{6}\ln z_{12}+\frac{\bar{c}}{6}\ln\bar{z}_{12}-\frac{1}{2}\ln\Big(\frac{k \ln z_{12}+\bar{k}\ln \bar{z}_{12}}{2\pi}\Big)-\frac{1}{2}\,.
\end{equation}\\
Also the \textit{fluctuation} entropy 
can be computed as,
\begin{equation} \label{fluc}
    S^{f}=-\int_{-\infty}^{\infty}P_{1}(N_{A})\ln{P_{1}(N_{A})}\,dN_{A}= +\frac{1}{2}\ln\Big(\frac{k \ln z_{12}+\bar{k}\ln \bar{z}_{12}}{2\pi}\Big)+\frac{1}{2}\,.
\end{equation}\\
Then we can write the total EE \cite{Bonsignori:2019naz} for Virasoro-Kac-Moody as,
\begin{equation} \label{VKM_consist}
    S_{vN}=S^{c}+S^{f}=\frac{c}{6}\ln z_{12}+\frac{\bar{c}}{6}\ln\bar{z}_{12}\,,
\end{equation}
which is the known result, made possible to obtain by the explicit cancellation as shown above. 

\subsection{Carrollian limit}
Now, we are in a position to take the Carrollian limit of (\ref{CFTKacMoody}) \footnote{We discuss the non-relativistic limit in Appendix~(\ref{AppC}).}.
As discussed, this is from the point of view of AdS radius $l$ blown up to infinity. The starting point towards this is to calculate the $l \rightarrow \infty$ Inönü-Wigner contraction of the Virasoro Kac-Moody algebra. As studied in detail \cite{Basu:2017aqn,Bagchi:2022xug}, this gives the BMS$_{3}$-Kac-Moody algebra. One could also have a look at Appendix \eqref{AppB} for a short introduction to the basics.
\medskip

A straightforward Carrollian limit of the EE, in \eqref{VKM_consist}, would give us that of a Carrollian theory whose bulk dual space-time is the null-orbifold (NO) solution, which plays the role of a flat space version of zero temperature BTZ black hole. As an aside, we note that, for a theory which lives on the boundary of the 
GFS, one needs to map the result \eqref{VKM_consist} to a cylinder and then take the limit \cite{Basu:2015evh}. The NO and the GFS solutions are, respectively, the special cases for $\mathcal{N} = 0, \mathcal{M} = 0$ and $\mathcal{N} = 0, \mathcal{M} = -1$ of the generic metric in asymptotically flat space-time:
\begin{eqnarray}\label{minkgen}
    ds^2 =  \mathcal{M}(\phi) du^2 - 2 du\,dr + 2 \mathcal{N} (u,\phi) du\, d\phi +r^2 d\phi^2;\qquad \, 2 \partial_u \mathcal{N} = \partial_\phi \mathcal{M}.
\end{eqnarray}

Without going to the holographic computation involving the Wilson lines for gauge fields valued in the BMS-Kac-Moody algebra, we therefore conjecturally propose the Carrollian limit of \eqref{CFTKacMoody} to be \textit{the SREE for the subsystem in the boundary of NO}. We do so directly using (\ref{param}), (\ref{param1}) and (\ref{param2}). The specific ultra-relativistic limit prescriptions then yield the following SREE of the ground state for BMS$_3$-Kac-Moody, 
\begin{align}
\begin{split} \label{BMSKacMoody}
S_{\textrm{BMSU(1)}}(N_A)= \frac{c_{L}}{6}\ln{x_{12}}+\frac{c_{M}}{6}\frac{t_{12}}{x_{12}}-\frac{1}{2}\ln\Big(\frac{k_{J}}{2\pi}\ln{x_{12}}+\frac{k_{P}}{2\pi}\frac{t_{12}}{x_{12}}\Big)-\frac{1}{2}\, .
\end{split}
\end{align}
Here $c_L$ and $c_M$ are defined in (\ref{BMSident}) after replacing $\bar{c}$ by $-\bar{c}\,$. Also, we need to replace $\bar{k}$ by $-\bar{k}\,$ to take a consistent limit.
These again follow from the choice of the representation of the BMS$_3$-Kac-Moody generators as discussed in Appendix~(\ref{AppB}). Also, while taking the Carrollian limit in the above, we made the following identifications as per \cite{Basu:2017aqn, Bagchi:2022xug} for the Chern-Simons levels:
\begin{equation}
k_{J}=k-\bar{k} \,,\quad k_{P}=\lim_{\beta \rightarrow \infty }\frac{1}{\beta}(k+\bar{k})\,.
\end{equation}
These $k_{J,P}$ are the levels for the two $U(1)$ currents in the BMS$_3$-Kac-Moody algebra, borne out of a contraction with the infinite boost parameter in \eqref{param2}.
For an equal-time interval on the boundary again $t_{12} = 0$ and for the bulk dual to be Einstein gravity we have $c_L = 0$. This means that the first two terms of \eqref{BMSKacMoody} vanish. However, one can still have both $k_J$ and $k_P$ non-zero at the cost of giving up chiral symmetry in the abelian Chern-Simons theory, in which case \eqref{BMSKacMoody} becomes:
\begin{eqnarray} \label{BMS_sym_res}
    S_{\textrm{BMSU(1)}}(N_A) = -\frac{1}{2}\ln \left(\frac{k_J}{2\pi} \ln L \right)+\mathcal{O}(1)
\end{eqnarray}
for a spatial interval of length $L$. However, this contributes to the configuration entropy only. Just as in the CFT case, one can calculate the fluctuation entropy from each charge sector as in the Virasoro-Kac-Moody case \eqref{fluc} to find the exact cancellation of the configuration and the fluctuations entropy contributions, so that we recover zero EE for the full system. To see that explicitly, following (\ref{Gaussian}) we can write $P_1(N_A)$ in the Carrollian limit as
\begin{equation}
    P_{1}(N_{A})_{\textrm{BMSU(1)}}=\sqrt{\frac{2\pi}{k_{J}\ln{x_{12}+k_{P}\frac{t_{12}}{x_{12}}}}} ~e^{-\frac{2\pi^{2}N_{A}^{2}}{k_{J}\ln{x_{12}+k_{p}\frac{t_{12}}{x_{12}}}}}\,.
\end{equation}
From (\ref{conf}), we can easily see that in the Carrollian limit the configurational entropy becomes,
\begin{equation} \label{confBMSKM}
    S^{c}_{\textrm{BMSU(1)}}=\frac{c_{L}}{6}\ln{x_{12}}+\frac{c_{M}}{6}\frac{t_{12}}{x_{12}}-\frac{1}{2}\ln\Big(\frac{k_{J}}{2\pi}\ln{x_{12}}+\frac{k_{P}}{2\pi}\frac{t_{12}}{x_{12}}\Big)-\frac{1}{2}\,.
\end{equation}
Also the fluctuation entropy becomes,
\begin{equation} \label{flucBMSKM}
    S^{f}_{\textrm{BMSU(1)}}=\frac{1}{2}\ln\Big(\frac{k_{J}}{2\pi}\ln{x_{12}}+\frac{k_{P}}{2\pi}\frac{t_{12}}{x_{12}}\Big)+\frac{1}{2}\,.
\end{equation}
Now we can easily see that sum of (\ref{confBMSKM}) and (\ref{flucBMSKM}) gives the total entanglement entropy $\frac{c_{L}}{6}\ln{x_{12}}+\frac{c_{M}}{6}\frac{t_{12}}{x_{12}}$ which vanishes for an equal-time interval in the boundary for Einstein gravity as mentioned above. 
So all in all, the general structure of symmetry resolved entropies work out perfectly even for a BMS-Kac-Moody case, giving us plenty of solid ground to explore these quantities even more closely for non-Lorentzian holography.

\section{Conclusions and Discussion}\label{sec8}

\subsection*{Summary}
Let us first start by summarizing our results. We started with a boosted interval in $1+1$d CFT and computed SRRE and SREE by introducing an Aharonov-Bohm flux on the replica manifold, which resulted in getting an extra term proportional to $(c-\bar{c})$. Then, we considered two extreme cases: non-relativistic and ultra-relativistic limits of the same. These two limits give us the results for SRRE and SREE for GCFT and CCFT, respectively. We identified the boost parameter (or the inverse of it)  itself for the GCFT and CCFT respectively, with the contraction parameter to obtain the required result. Furthermore, for GCFT, we have also provided an intrinsic computation that matches the results from the limiting procedure. A derivation showing that the composite twist operators for GCFT also behave as primary operators has been provided, generalizing the derivation of \cite{Basu:2015evh,Basu:2017aqn}. This further provides a consistency check of our computation. Remarkably both for GCFT and CCFT, the so-called ``equipartition of entanglement'' holds, thereby generalizing it for the non-Lorentzian case. We have also discussed the effect of boost for individual charge sectors. Furthermore, we have extended our computation for system at finite temperature and finite spatial extent. We checked that the equipartition still holds. 
\medskip

An interesting observation that came out from our analysis is that, for an equal-time interval SREE can be non-zero but when we sum over all the charge sectors in the Carrollian theory, we get the total EE which vanishes for the same interval when $c_L=0.$ Furthermore, considering a possible connection with flat-space holography, we extended our analysis for the BMS$_3$-Kac-Moody case which includes two $U(1)$ charges, like the Virasoro cousin thereof, and commented on the connection with dual quantities in the bulk. For this case, we have also explicitly shown that the equipartition of entanglement holds in general. To the best of our knowledge, these are some of the first computations of charged Renyi (as well as entanglement) entropy for non-Lorentzian field theories.  

\subsection*{Future directions}
Several interesting avenues are worth exploring in the near future. First and foremost, we have considered a constant boost throughout our study. This corresponds to the action of the generator $x \partial_t$ in the Carrollian context. Keeping in mind the holographic correspondence, one can map one solution of flat-space gravity to another one by non-trivial supertranslation (equivalently thought of as a non-uniform boost). For example, solutions like flat-space cosmology can be generated from the global flat space by a non-trivial supertranslation charge. It has been confirmed that the stringent condition of strong-subadditivity is saturated by the EE for duals of global flat space but is violated by those of flat space cosmology solutions \cite{Apolo:2020bld}. It is, therefore imperative that we understand this violation of strong subadditivity condition in terms of more fine-grained quantity like SREE both from a field theory perspective as well as holographically. We hope to report this in near future.
\medskip

It will also be interesting to generalize our result for the third kind of generic asymptotically flat metric in \eqref{minkgen}, known as Flat Space Cosmology (FSC). These shifted-boost orbifolds of Minkowski spacetimes occur when we take $\mathcal{M}>0$, and are shown to be dual to a thermal CCFT. We can hope to do this either by using the limiting procedure \cite{Basu:2015evh,Basu:2017aqn}, or using an intrinsic Rindler computation similar to \cite{Jiang:2017ecm}. In general, it would be good to have a first-principle intrinsic computation of the SRRE for CCFT, which was beyond the scope of the current work. Furthermore, in this paper, we have considered the BMS$_3$-Kac-Moody theory with Kac-Moody currents coming from abelian $U(1)$ gauge fields. One can try to generalize it for non-abelian currents in Carrollian Kac-Moody systems similar to the Lorentzian WZW model setup in \cite{Calabrese:2021wvi}. 
\medskip

Since a large portion of our attention here was devoted to Carrollian theories, the study of which is merely nascent at this stage, there's a plethora of question one may be able to answer using SREE computations for this case. For example, everything we calculated in this manuscript is based on the highest weight vacuum for a CCFT (see Appendix \eqref{AppB} for some comments). However a true ultra-relativistic limit of Virasoro highest weight representation give rise to the so-called Induced representations \cite{Oblakthesis} of BMS algebra. It would be interesting to uncover the entanglement signatures of the theory built on this other vacuum in some future discussion. 
\medskip

In recent times, it has been observed that the Carrollian symmetry emerges in the dispersion-less limit in certain fermionic chain models with flat bands, and in the context of magic superconductivity of bilayer graphene \cite{Bagchi:2022eui,Ara:2023pnn}. This provides us with ample opportunity to set up our analysis (numerically) for spin-chain models with flat bands and to compare numerical results with those of the analytical results presented in this paper. We again hope to report on this issue in the near future. Moreover, since we conclusively showed the presence of equipartition of SREE even for non-Lorentzian cousins of CFT, it would be interesting to see whether its saturation \cite{Castro-Alvaredo:2024azg} in massive Carrollian QFT holds as well. Last but not least, it will be interesting to generalize our results for other measures of entanglement as well as for complexity by generalizing the results of \cite{Bhattacharyya:2023sjr} for BMS$_3$-Kac-Moody. 

\section*{Acknowledgments}
It is a pleasure to thank Arjun Bagchi for discussions and comments on the draft.
Arpan Bhattacharyya (AB) thanks the speakers and participants of the workshop ``Quantum Information in QFT and AdS/CFT-III" organized at IIT Hyderabad between 16-18th September, 2022 and funded by SERB through a Seminar Symposia (SSY) grant (SSY/2022/000446), ``Quantum Information Theory in Quantum Field Theory and Cosmology" held in 4-9th June, 2023 hosted by Banff International Research Centre at Canada and ``Holography, Strings and other fun things" at IIT Kanpur and BITS Pilani (Goa Campus) between 19-23th February, 2024.  AB and Aritra Banerjee (ArB) would also like to thank the Department of Physics of BITS Pilani, Goa Campus for hospitality during the course of this work. ArB is supported in part by an OPERA grant and a seed grant NFSG/PIL/2023/P3816 from BITS-Pilani. The grants support the research of RB are: CRG/2020/002035, MTR/ 2022/000795 from SERB, India, DST/IC/Austria/P-9/2021 from DST, India and OeAD Austria; and OPERA grant from BITS Pilani. AB is supported by the Core Research Grant (CRG/2023/ 001120), Mathematical Research Impact Centric Support Grant (MTR/2021/ 000490) by the Department of Science and Technology Science and Engineering Research Board (India), India and Relevant Research Project grant (202011BRE03RP06633-BRNS) by the Board of Research in Nuclear Sciences (BRNS), Department of Atomic Energy (DAE). AB also acknowledges the associateship program of the Indian Academy of Science, Bengaluru. Nilachal Chakrabarti (NC) is supported by the Director's Fellowship of the Indian Institute of Technology Gandhinagar.

\appendix

\section{Representations of GCA/CCA} 
\label{AppB}
\subsection*{Two dimensional contractions}
Let us mention some basic details about two dimensional Galilean and Carrollian Conformal Algebra (GCA and CCA) in this appendix. For details, we encourage the reader to refer to \cite{Bagchi:2009pe,Bagchi:2019xfx}, and references therein. The generators of the GCA in $d=2$ involves the $L_n$'s and the $M_n$'s, with which the centrally extended version of the GCA is given by:
\begin{eqnarray}\label{gca2d}
&&[L_n, L_m] = (n-m) L_{n+m} + \frac{c_L}{12}(n^3-n) \delta_{n+m,0} \, , \cr
&&[L_n, M_m] = (n-m) M_{n+m} + \frac{c_M}{12}(n^3-n) \delta_{n+m,0} \, , \quad [M_n, M_m]= 0. 
\end{eqnarray}
The infinite dimensional 2d GCA can be obtained as a contraction of two copies of the Virasoro algebra. If we define the relativistic conformal algebra generators by $\L_n, \bL_n$, the following combination of the Virasoro generators give rise to the 2d GCA in the limit $\epsilon \to 0$:
\begin{equation}\label{nr-lim}
\L_n + \bL_n = L_n, \quad \L_n - \bL_n = \frac{1}{\epsilon} M_n
\end{equation}
The central terms $c_L, c_M$ in the quantum 2d GCA \eqref{gca2d} can be linked to the original CFT central terms in this non-relativistic limit:
\begin{equation}\label{c-nr}
c+ \bar{c} = c_L, \quad c- \bar{c} = \frac{1}{\epsilon} c_M
\end{equation}
It is instructive to note that by looking at the representations of the Virasoro generators on the complex plane, viz. 
\begin{equation}\label{genz}
\L_n = z^{n+1} \partial_z, \quad \bL_n = \z^{n+1} \partial_\z
\end{equation}
and taking $z = t + \epsilon x, \, \z = t -\epsilon x$, one can readily obtain the 2d version of the GCA generators in the limit $\epsilon \to 0$. 
However, there exists another contraction of the Virasoro algebra which generates the 2d Carroll Conformal Algebra (CCA):
\begin{equation}\label{ur-lim}
\L_n - \bL_{-n} = L_n, \quad \L_n + \bL_{-n} = \frac{1}{\epsilon} M_n
\end{equation}
Intriguingly, this contraction also generates the same algebra in two dimensions as the GCA.
 The interpretation of this limit is reversed from the previous one, i.e. the role of $x$ and $t$ are interchanged in \eqref{genz}. This is the limit where the speed of light tends to zero and is hence the ultra-relativistic limit of the conformal algebra. The mapping from the 2d CFT also changes the interpretation of the central terms in this limit. In this case, the central terms are given by 
\begin{equation}\label{c-ur}
c_L = c - \bar{c}, \quad c_M = \epsilon (c + \bar{c}) 
\end{equation}
which is to be contrasted to (\ref{c-nr}). Although classically equivalent, the representation theory for the two algebras in the quantum regime differs by a lot. Consider zero modes of the generators $L,M$ acting on a primary state having weights $h_L, h_M$, then we get:

\begin{equation}
L_0 |h_L, h_M \rangle = h_L |h_L, h_M \rangle, \quad M_0 |h_L, h_M \rangle = h_M |h_L, h_M \rangle
\end{equation}
Using the two different limits, we see that the contracted weights in the non-Lorentzian theory can be expressed in terms of the Virasoro weights $h, \h$ (where the Virasoro states are labeled as $\L_0  |h, \h\rangle= h |h, \h\rangle, \, \bL_0 |h, \h\rangle= \h |h, \h\rangle$) as:
\begin{equation}\label{hurnr}
\mbox{Galilean limit:} \quad h+\h = h_L, \, h- \h = \frac{1}{\epsilon} h_M; \quad \mbox{Carrollian limit:} \quad h-\h = h_L, \, h+\h = \frac{1}{\epsilon} h_M
\end{equation}

Now a comparison between \eqref{c-nr}, \eqref{c-ur} and \eqref{hurnr} clearly explains the replacement of $\bar{c}$ with $-\bar{c}$ and $\bar{\Delta}$ to $-\bar{\Delta}$ in the main text after taking the ultra-relativistic limits. In fact, this can be explained by a curious inner automorphism associated to the parent Virasoro algebra, where one can replace:
\begin{equation}
    \mathcal{\bar{L}}_n \to \mathcal{-\bar{L}}_{-n}, ~~\bar{c}\to -\bar{c},
\end{equation}
which keeps the centrally extended algebra invariant. One can see that this creates the perceived duality between UR and NR contractions discussed above. In fact taking this flipped representation in conjunction with a ultra-relativistic limit directly lands one on the BMS highest weight representation. 

\subsection*{Representation of $\textrm{BMS}_3 \oplus \mathfrak{u}(1)_{k_J} \oplus \mathfrak{u}(1)_{k_P}: $}
These algebras can be extended when there are $U(1)$ currents present \cite{Bagchi:2022xug}. If $\mathcal{J}_n$ and $\bar{\mathcal{J}}_n$ are two $U(1)$ generators and $k$ and $\bar{k}$ are the levels of the current generators, then from the  algebra of $\mathfrak{vir} \oplus \mathfrak{u}(1)_k \oplus \mathfrak{u}(1)_{\bar{k}}\,,$
\begin{align}
\begin{split}
&[\mathcal{L}_n,\mathcal{L}_m]=(n-m)\mathcal{L}_{n+m}+\frac{c}{12}(n^3-n)\delta_{n+m,0},\quad [\mathcal{L}_n,\mathcal{J}_m]=-m\,\mathcal{J}_{n+m},\quad [\mathcal{J}_n,\mathcal{J}_m]=k\,n\,\,\delta_{n+m,0},\\&
[\mathcal{\bar{L}}_n,\mathcal{\bar{L}}_m]=(n-m)\mathcal{L}_{n+m}+\frac{\bar{c}}{12}(n^3-n)\delta_{n+m,0},\quad [\mathcal{\bar{L}}_n,\mathcal{\bar{J}}_m]=-m\,\mathcal{\bar{J}}_{n+m},\quad [\mathcal{\bar{J}}_n,\mathcal{\bar{J}}_m]=\bar{k}\,n\,\,\delta_{n+m,0}\,,
\end{split}
\end{align}
We get the following BMS-Kac-Moody algebra after taking the Carrollian limit \cite{Bagchi:2022xug}, 
\begin{align}
\begin{split} \label{BMSKac}
&[L_n, L_m] = (n-m) L_{n+m} + \frac{c_L}{12}(n^3-n) \delta_{n+m,0}\,,\\&
[L_n, M_m] = (n-m) M_{n+m} + \frac{c_M}{12}(n^3-n) \delta_{n+m,0} \,,\\&[L_n, J_m]= -m\,J_{n+m}\,,[L_n,P_m]=-m\,P_{n+m}\,,[M_n,J_m]=-m\,P_{n+m}\,,\\&
[J_n,J_m]=k_J\,n\,\delta_{n+m,0}\,,[J_n,P_m]=k_P\,n\,\delta_{n+m,0}\,.
\end{split}
\end{align}
Here the contracted generators, 
\begin{align}
\begin{split}\label{contrac1}
\L_n - \bL_{-n} = L_n, \quad \L_n + \bL_{-n} = \frac{1}{\epsilon} M_n\,\quad \mathcal{J}_n-\mathcal{\bar{J}}_{-n}=J_n,\quad \mathcal{J}_n+\mathcal{\bar{J}}_{-n}=\frac{1}{\epsilon}P_n
\end{split}
\end{align}
and the current levels: $$k_J=k-\bar{k}\, \quad k_p=\epsilon(k+\bar{k})\,.$$
Like before, one can also define the following Galilean contraction for the same generators: 
\begin{align}
\begin{split} \label{contrac}
\L_n + \bL_n = L_n, \quad \L_n - \bL_n = \frac{1}{\epsilon} M_n\,\quad \mathcal{J}_n+\mathcal{\bar{J}}_{n}=J_n,\quad \mathcal{J}_n-\mathcal{\bar{J}}_{n}=\frac{1}{\epsilon}P_n
\end{split}
\end{align}
and $$k_J=k+\bar{k}\, \quad k_p=\epsilon(k-\bar{k})\,.$$ 
These generators also satisfy the same non-Lorentzian algebra as mentioned in (\ref{BMSKac}). But the contraction mentioned in (\ref{contrac}) should be contrasted to (\ref{contrac1}) like before. As discussed earlier,  for this case there is also a duality between ultra-relativistic and non-relativistic contractions \cite{Bagchi:2022xug}.
\begin{equation}
    \mathcal{\bar{L}}_n \to \mathcal{-\bar{L}}_{-n}, \quad \mathcal{\bar{J}}_n \to \mathcal{-\bar{J}}_{-n},\quad  \bar{c}\to -\bar{c},\quad \bar{k} \to -\bar{k}\,.
\end{equation}
Hence our choices while taking limits in the main text all come together nicely.

\section{One point functions: A consistency check} \label{AppA}

We can arrive at (\ref{Final}) by taking the appropriate limit of the CFT result given in \cite{Goldstein:2017bua}. We start with the following CFT result, 
\begin{align}
\begin{split} \label{check}
&\langle T(w')\rangle_{\mathcal{R}_{n,\alpha}}=\Big[\frac{\Delta}{n^2}+\Big(1-\frac{1}{n^2}\Big)\frac{c}{24}\Big]\frac{(z_1-z_2)^2}{(w'-z_1)(w'-z_2)}\,,\\& \langle \bar{T}(\bar{w}')\rangle_{\mathcal{R}_{n,\alpha}}=\Big[\frac{\bar{\Delta}}{n^2}+\Big(1-\frac{1}{n^2}\Big)\frac{\bar{c}}{24}\Big]\frac{(\bar{z}_1-\bar{z} _2)^2}{(\bar{w}'-\bar{z}_1)(\bar{w}'-\bar{z}_2)}\,.
\end{split}
\end{align}
Then we use the following identifications:
\begin{align}
\begin{split}
&\langle T_1(t',x')\rangle_{\mathcal{R}_{n,\alpha}}=\lim_{\epsilon\rightarrow 0}\Big(\langle T(w')\rangle_{\mathcal{R}_{n,\alpha}}+\langle \bar{T}(\bar{w}')\rangle_{\mathcal{R}_{n,\alpha}}\Big)\,,\\&\langle T(t',x')\rangle_{\mathcal{R}_{n,\alpha}}=\lim_{\epsilon\rightarrow 0}\epsilon\Big(\langle T(w')\rangle_{\mathcal{R}_{n,\alpha}}-\langle \bar{T}(\bar{w}')\rangle_{\mathcal{R}_{n,\alpha}}\Big)
\end{split}
\end{align}
and 
\begin{align}
\begin{split}
w'=t'+\epsilon\, x', \bar{w}'=t'-\epsilon x',\,\quad   z_i=t_i+\epsilon\, x_i, \bar{z}_i=t_i-\epsilon\,x_i,~~ i=1,2\,. 
\end{split}
\end{align}
From (\ref{limit}) it can be easily identified that this $\epsilon$ plays the role of $\beta.$ Then expanding (\ref{check}) and using (\ref{comapre1}) we can arrive at (\ref{Final}). This serves as yet another consistency check of our computation.\\

\section{Non-relativistic limit of SREE in Virasoro-Kac-Moody}\label{AppC}

 We consider take the non-relativistic limit of (\ref{CFTKacMoody}) to get the corresponding result of the symmetry-resolved entropy for GCFT-Kac-Moody. We follow the procedure mentioned in Sec.~(\ref{sec4}). Then we get 
\begin{align}
\begin{split} \label{GCFTKacMoody}
S_{\textrm{GCFTU(1)}}(N_A)= \frac{c_{L}}{6}\ln{t_{12}}+\frac{c_{M}}{6}\frac{x_{12}}{t_{12}}-\frac{1}{2}\ln\Big(\frac{k_{J}}{2\pi}\ln{t_{12}}+\frac{k_{P}}{2\pi}\frac{x_{12}}{t_{12}}\Big)\,,
\end{split}
\end{align}
where $c_L$ and $c_M$ are defined in (\ref{comapre1}) and following \cite{Bagchi:2022xug} we can identify, 
\begin{align}
k_J=k+\bar{k},\quad k_P=\beta (k-\bar{k})\,.
\end{align}
Just like in the Carrollian case, we can show that the sum of configurational and fluctuation entropy gives the total entanglement entropy. 
\bibliographystyle{JHEP}
\bibliography{biblio}

\end{document}